\documentclass[useAMS,usenatbib]{mn2e}

\usepackage[dvips]{graphicx}

\usepackage{color}
\usepackage{times,ulem}
\newcommand{\aj}{AJ}
\newcommand{\apj}{ApJ}
\newcommand{\apjs}{ApJS}
\newcommand{\aap}{A\&A}
\newcommand{\aaps}{A\&AS}
\newcommand{\mnras}{MNRAS}

\title[Statistical tests in variability of AGNs]{Microvariability in AGNs:
  study of different statistical methods - I. Observational analysis}
\author[L. Zibecchi et al.]{L.~Zibecchi,$^{1,2,\star}$
  I.~Andruchow,$^{1,2}$
  S. A.~Cellone,$^{1,2}$ D. D.~Carpintero,$^{1,2}$ \newauthor G. E.~Romero$^{1,3}$ and
  J. A.~Combi$^{1,3}$\\ 
$^{1}$ Facultad de Ciencias Astron\'omicas y Geof\'isicas, Universidad Nacional de La Plata,
  Paseo del Bosque, B1900FWA La Plata, Argentina.\\ 
$^{2}$ Instituto de Astrof\'isica La Plata (IALP), CONICET-UNLP, Paseo del Bosque, B1900FWA La Plata, Argentina.\\ 
$^3$ Instituto Argentino de Radioastronom\'ia (IAR), CONICET, C.C. 5,
  1894 Villa Elisa, Argentina.\\ 
$^{\star}$ {\it Contact e-mail: lzibecchi@fcaglp.unlp.edu.ar}}

\begin{document}

\date{Accepted 2017 January 7. Received 2017 January 7; in original form 2016 September 23}

\pagerange{\pageref{firstpage}--\pageref{lastpage}} \pubyear{2017}

\maketitle

\label{firstpage}

\begin{abstract}
  We present the results of a study of different statistical
  methods currently used in the literature to analyse the
  (micro)variability of active galactic nuclei (AGNs) from ground-based
  optical observations. In particular, we focus on the comparison
  between the results obtained by applying the so-called $C$ and $F$ statistics,
  which are based on the ratio of standard deviations and variances,
  respectively. The motivation for this is that the implementation of these
  methods leads to different and contradictory results, making the
  variability classification of the light curves of a certain source
  dependent on the statistics implemented.

  For this purpose, we re-analyse the results on an AGN sample
  observed along several sessions with the 2.15m \lq Jorge
  Sahade\rq~telescope ({\sc casleo}), San Juan, Argentina.  For each
  AGN we constructed the nightly differential light curves. We thus
  obtained a total of 78 light curves for 39 AGNs, and we then applied
  the statistical tests mentioned above, in order to re-classify the
  variability state of these light curves and in an attempt to find
  the suitable statistical methodology to study photometric
  (micro)variations. We conclude that, although the $C$ criterion is not
  proper a statistical test, it could still be a suitable parameter to
  detect variability and that its application allows us to get more
  reliable variability results, in contrast with the $F$ test.
\end{abstract}

\begin{keywords}
methods: statistical -- galaxies: active -- techniques: photometric.
\end{keywords}

\section{Introduction}
\label{sec:intro}

Active galactic nuclei (AGNs) are well known for their extreme
electromagnetic emission (reaching values of radiating powers up to
10$^{46}$ \textrm{erg s$^{-1}$}), which is spread over the whole
spectrum (from radio to X-rays bands). This emission presents, in some
cases, a peak in the UV region and significant emission in the X-rays
and infrared bands.

Most AGNs, and blazars in particular, are characterized by
va\-ria\-bi\-li\-ty in their optical flux. The time-scales of these
changes span a range from days to years, but variations on time-scales
of hours or minutes also take place. This latter phenomenon is known
as \textit{microvariability}, and it has been studied and reported by
several authors in the last decades
(e.g. \citealp*{1989Natur.337..627M, 1990AJ....100..347C,
  2000A&A...360L..47R}; \citealp{2011MNRAS.412.2717J}).
Microvariability studies provide important information about size
limits for the emitting regions and can provide constraints on
different models of the electromagnetic emission. However, spurious
variability results may be obtained due to: (i) systematic
errors introduced by contamination from the host galaxy light
\citep*{2000AJ....119.1534C}; (ii) i\-nap\-pro\-pria\-te
observing/photometric methodologies \citep*{2007MNRAS.374..357C}, and
(iii) the inadequate use of statistical methods for the
detection of variability
\citep{2010AJ....139.1269D,2011MNRAS.412.2717J}.

In the present work, we focus on the last item. In the literature, we
may find a great diversity of statistical tests used to assess the
significance of variability results. The most commonly used are: the
$\chi^2$ test, which compares a sample variance of the possibly
variable target with a theoretically calculated variance for a
non-variable object, proposed by \citet*{1976AJ.....81..919K}, and
used both for photometric and polarimetric time series
(\citealp*{1994A&A...288..731R}; \citealp{2003A&A...409..857A,
  2005A&A...442...97A, 2010AJ....139.1269D}); the one way analysis of
variance (ANOVA), which is a family of tests that compare the means of a
number of samples \citep{1998ApJ...501...69D,
  2004A&A...421...83R,2009AJ....138..991R, 2010AJ....139.1269D}; the
$C$ criterion, which involves the ratio of standard deviations of two
distributions (\citealp*{1988AJ.....95..247H};
\citealp{1999A&AS..135..477R,2002A&A...390..431R};
\citealp*{2005A&A...442...97A, 2010AJ....139.1269D};
\citealp{2011MNRAS.412.2717J, 2011BAAA...54..325Z}); and the $F$ test,
which takes into account the ratio be\-tween the variances of two
distributions \citep{2010AJ....139.1269D, 2011MNRAS.412.2717J}.

Contradictory and diverse results are usually obtained from these
statistics, and it is of course desirable that the
classification of the state of variability of a certain source should
be independent from the statistical method used. In order to find the
most reliable test to study variability, we took advantage of a
significantly large data set of AGN microvariability observations
obtained with the same instrumental setup and reduced in a
homogeneous way.

In Section~\ref{sec:odr}, we present the sample of AGNs and the method
to generate the differential light curves (DLCs). In Section~\ref{sec:stsv},
we describe the $C$ and $F$ statistics, respectively, and we present
our results, ma\-king a comparison between tests. In
Section~\ref{sec:iicc}, we make a deeper study on the $C$
criterion. In Section~\ref{sec:discu}, we present the results of the
implementation of both statistics to the field stars, and finally, in
Section~\ref{sec:summ} we discuss the results found and summarize our
conclusions. Appendix~\ref{apendice} describes in detail the $D$ test
mentioned in Section 4.1.

\section{Observations and data reduction}
\label{sec:odr}

We worked with a sample of 23 southern AGNs reported in
\citet{1999A&AS..135..477R}, and 20 {\sc egret} blazars, studied by
\citet{2002A&A...390..431R}. The data in both papers were based on
observations taken with the 2.15m \lq Jorge Sahade\rq~telescope, {\sc
  casleo}, Argentina, between 1997 April and 2001 July. The telescope
was equipped with a {\it liquid-nitrogen-cooled} CCD camera, using a
Tek-1024 chip with a gain of 1.98 electrons/adu and a {\it
  read-out noise} of 9.6 electrons. A {\it focal-reducer} providing a
scale of 0.813 arcsec pixel$^{-1}$ was also used. Since three sources are
repeated in both samples, and the object PKS\,1519$-$273 was excluded
because the original data could not be recovered, we have studied a
total sample of 39 AGN.

In the original publications, objects were classified as:
\textit{quasars} (QSO), within which there are the \lq
\textit{radioquiet}\rq~(RQQ) and \lq \textit{radio\-loud}\rq~(RLQ);
and \textit{BL\,Lac objects}, which have been categorised in \lq
\textit{radio-selected}\rq~(RBL) and in \lq
\textit{X-rays-selected}\rq~ (XBL). After se\-ve\-ral revisions, and
following the publication of the first ca\-talogue of the satellite
instrument {\it Fermi-}LAT (Large Area Teles\-cope; \citet{LAT-AGN}),
the blazars are now broadly divided into BL\,Lacs and flat-spectrum
radio quasars (FSRQ), and further sub-classified based on the
frequency at which the synchrotron peak of the spectral energy
distribution falls, as: \textit{low synchrotron peak}, LSP blazars,
\textit{intermediate synchrotron peak}, ISP blazars, and \textit{high
  synchrotron peak}, HSP blazars \citep{LAT-AGN}.

The sample of AGNs is presented in Table~\ref{tab:obj}, where we give
the name of the source, type of AGN, right ascension ($\alpha$),
declination ($\delta$), redshift ($z$) and the visual magnitude
($m$). These values were taken from the NASA/IPAC Extragalactic
  Database\footnote{http://ned.ipac.caltech.edu/} and from the
references cited in the table. Observations are charac\-terized by
seeing values between 2.0 and $\ga$ 4.0 \textrm{arcsec}, exposure
times ranging between 2 and 15 \textrm{min}, and airmass values
between 1.00 and 2.40.

\begin{table*}
\begin{minipage}{100mm}
\caption[Data for the objects]{Data for the objects. $\bullet$ \citet{2015ApJ...810...14A}; 
$\star$ \citet{VCV-2010}; $\ast$ \citet{CNTC-2007}; $\dagger$ \citet{2011ApJS..194...29R}.}
\label{tab:obj}
\begin{tabular}{rccccc}
 \hline \noalign{\smallskip}
Object & Type & $\alpha$ (J2000.0) & $\delta$ (J2000.0) & $z$ & $m$\\
 & & h m s & $\degr \quad \arcmin \quad \arcsec$ & & Visual mag.\\
\noalign{\smallskip} \hline \noalign{\smallskip}
0208$-$512 & BLL/LSP$^{\bullet}$ & 02:10:46 & $-$51:01:02 & 1.003 & 16.9\phantom{0}\\
0235$+$164 & BLL/LSP$^{\bullet}$ & 02:38:39 & $+$16:36:59 & 0.904 & 18.0\phantom{0}\\
0521$-$365 & BLL/LSP$^{\bullet}$ & 05:22:58 & $-$36:27:31 & 0.55\phantom{0} & 14.5\phantom{0}\\
0537$-$441 & BLL/LSP$^{\bullet}$ & 05:38:50 & $-$44:05:09 & 0.894 & 15.5\phantom{0}\\
0637$-$752 & FSRQ/LSP$^{\bullet}$ & 06:35:47 & $-$75:16:17 & 0.651 & 15.75\\
1034$-$293 & QSO$^{\star}$ & 10:37:16 & $-$29:34:03 & 0.312 & 16.46\\
1101$-$232 & BLL/HSP$^{\bullet}$ & 11:03:38 & $-$23:29:31 & 0.186 & 16.55\\
1120$-$272 & QSO$^{\ast}$ & 11:23:02 & $-$27:30:04 & 0.389 & 16.8\phantom{0}\\
1125$-$305 & QSO$^{\ast}$ & 11:27:32 & $-$30:44:46 & 0.673 & 16.3\phantom{0}\\
1127$-$145 & FSRQ/LSP$^{\bullet}$ & 11:30:07 & $-$14:49:27 & 1.187 & 16.9\phantom{0}\\
1144$-$379 & FSRQ/LSP$^{\bullet}$ & 11:47:01 & $-$38:12:11 & 1.048 & 16.2\phantom{0}\\
1157$-$299 & QSO$^{\ast}$ & 11:59:43 & $-$30:11:53 & 0.207 & 16.4\phantom{0}\\
1226$+$023 & FSRQ/LSP$^{\bullet}$ & 12:29:07 & $+$02:03:08 & 0.158 & 12.86\\
1229$-$021 & QSO$^{\star}$ & 12:32:00 & $-$02:24:05 & 1.045 & 17.7\phantom{0}\\
1243$-$072 & QSO$^{\star}$ & 12:46:04 & $-$07:30:47 & 1.286 & 19.0\phantom{0}\\
1244$-$255 & FSRQ/LSP$^{\bullet}$ & 12:46:47 & $-$25:47:49 & 0.638 & 17.41\\
1253$-$055 & FSRQ/LSP$^{\bullet}$ & 12:56:11 & $-$05:47:22 & 0.536 & 17.75\\
1256$-$229 & QSO$^{\star}$ & 12:59:08 & $-$23:10:39 & 0.481 & 17.3\phantom{0}\\
1331$+$170 & FSRQ$^{\dagger}$ & 13:33:36 & $+$16:49:04 & 2.084 & 16.71\\
1334$-$127 & FSRQ/LSP$^{\bullet}$ & 13:37:40 & $-$12:57:25 & 0.539 & 17.2\phantom{0}\\
1349$-$439 & BLL/LSP$^{\bullet}$ & 13:52:57 & $-$44:12:40 & 0.05\phantom{0} & 16.37\\
1424$-$418 & FSRQ/LSP$^{\bullet}$ & 14:27:56 & $-$42:06:19 & 1.522 & 17.7\phantom{0}\\
1510$-$089 & FSRQ/LSP$^{\bullet}$ & 15:12:50 & $-$09:06:00 & 0.361 & 16.5\phantom{0}\\
1606$+$106 & FSRQ/LSP$^{\bullet}$ & 16:08:46 & $+$10:29:08 & 1.226 & 18.5\phantom{0}\\
1622$-$297 & FSRQ/LSP $^{\bullet}$ & 16:26:06 & $-$29:51:27 & 0.815 & 20.5\phantom{0}\\
1741$-$038 & QSO$^{\star}$ & 17:43:59 & $-$03:50:05 & 1.054 & 18.6\phantom{0}\\
1933$-$400 & FSRQ/LSP$^{\bullet}$ & 19:37:16 & $-$39:58:02 & 0.965 & 18.0\phantom{0}\\
2005$-$489 & BLL/HSP$^{\bullet}$ & 20:09:25 & $-$48:49:54 & 0.071 & 13.4\phantom{0}\\
2022$-$077 & FSRQ/LSP$^{\bullet}$ & 20:25:41 & $-$07:35:53 & 1.388 & 18.5\phantom{0}\\
2155$-$304 & BLL/HSP$^{\bullet}$ & 21:58:52 & $-$30:13:32 & 0.116 & 13.1\phantom{0}\\
2200$-$181 & QSO$^{\ast}$ & 22:03:12 & $-$18:01:43 & 1.16\phantom{0} & 15.3\phantom{0}\\
2230$+$114 & FSRQ/LSP$^{\bullet}$ & 22:32:36 & $+$11:43:51 & 1.037 & 17.33\\
2254$-$204 & BLL/LSP$^{\bullet}$ & 22:56:41 & $-$20:11:41 & ... & 16.6\phantom{0}\\
2316$-$423 & BLL/HSP$^{\bullet}$ & 23:19:06 & $-$42:06:49 & 0.054 & 16.0\phantom{0}\\
2320$-$035 & FSRQ/LSP$^{\bullet}$ & 23:23:32 & $-$03:17:05 & 1.41\phantom{0} & 18.6\phantom{0}\\
2340$-$469 & QSO$^{\ast}$ & 23:43:14 & $-$46:40:03 & 1.97\phantom{0} & 16.4\phantom{0}\\
2341$-$444 & QSO$^{\ast}$ & 23:43:47 & $-$44:07:19 & 1.9\phantom{0}\phantom{0} & 16.5\phantom{0}\\
2344$-$465 & QSO$^{\ast}$ & 23:46:41 & $-$46:12:30 & 1.89\phantom{0} & 16.4\phantom{0}\\
2347$-$437 & QSO$^{\ast}$ & 23:50:34 & $-$43:26:00 & 2.885 & 16.3\phantom{0}\\
\noalign{\smallskip}\hline
\end{tabular}
\end{minipage}
\end{table*}

\subsection{Differential photometry}
\label{sec:dph}
 
The statistical analysis is made on DLCs. These curves are obtained by
applying standard differential photometry techniques, as were
developed by \citet{1986PASP...98..802H}. The observations involve
repeated short exposures of a certain field that contains the source
of interest.  Other stars in the frame are used for comparison and
control in the reduction process, which results in instrumental
magnitudes of all the objects. The principal advantage of differential
photometry is that there is no need for perfect photometric
nights. Following \citet{1986PASP...98..802H}, the source of interest
is designed by V, and a comparison and a control stars by C and K,
respectively.  It is important to highlight that both stars should not
be variable.
 
With the instrumental magnitudes, $m_V - m_C$ and $m_K - m_C$ are
calculated, being the last one important because (i) variability in
the comparison and/or control star can be detected; (ii) intrinsic
instrumental precision is measured, and (iii) it provides a
com\-pa\-ri\-son to determine whether the light curve of the source is
variable or not.

Several objects of the sample have been observed along more than one
night, making a total of 78 data sets (i.e. each data set corresponds
to observations taken along one night for a given object). For each
set, we generated a DLC, using the software {\sc
  iraf}\footnote{{\sc iraf} is distributed by the National Optical
  Astronomy Observatories, which are operated by the Association of
  Universities for Research in Astronomy, Inc., under cooperative
  agreement with the National Science Foundation.}  (Image Reduction
and Analysis Facility). For the photometry, we used an optimal
aperture radius, which is determined taking into account the apparent
size and the brightness of the host galaxy, when appropriate
\citep{2000AJ....119.1534C}.  For almost all the AGNs in the sample,
we took the same radius of 6.5~\textrm{arcsec}, except for
PKS\,1622$-$297 for which we used a radius of 3.5~\textrm{arcsec}
because the field of this object is particularly crowded.

\begin{figure}
\centering
\includegraphics[width=0.45\textwidth]{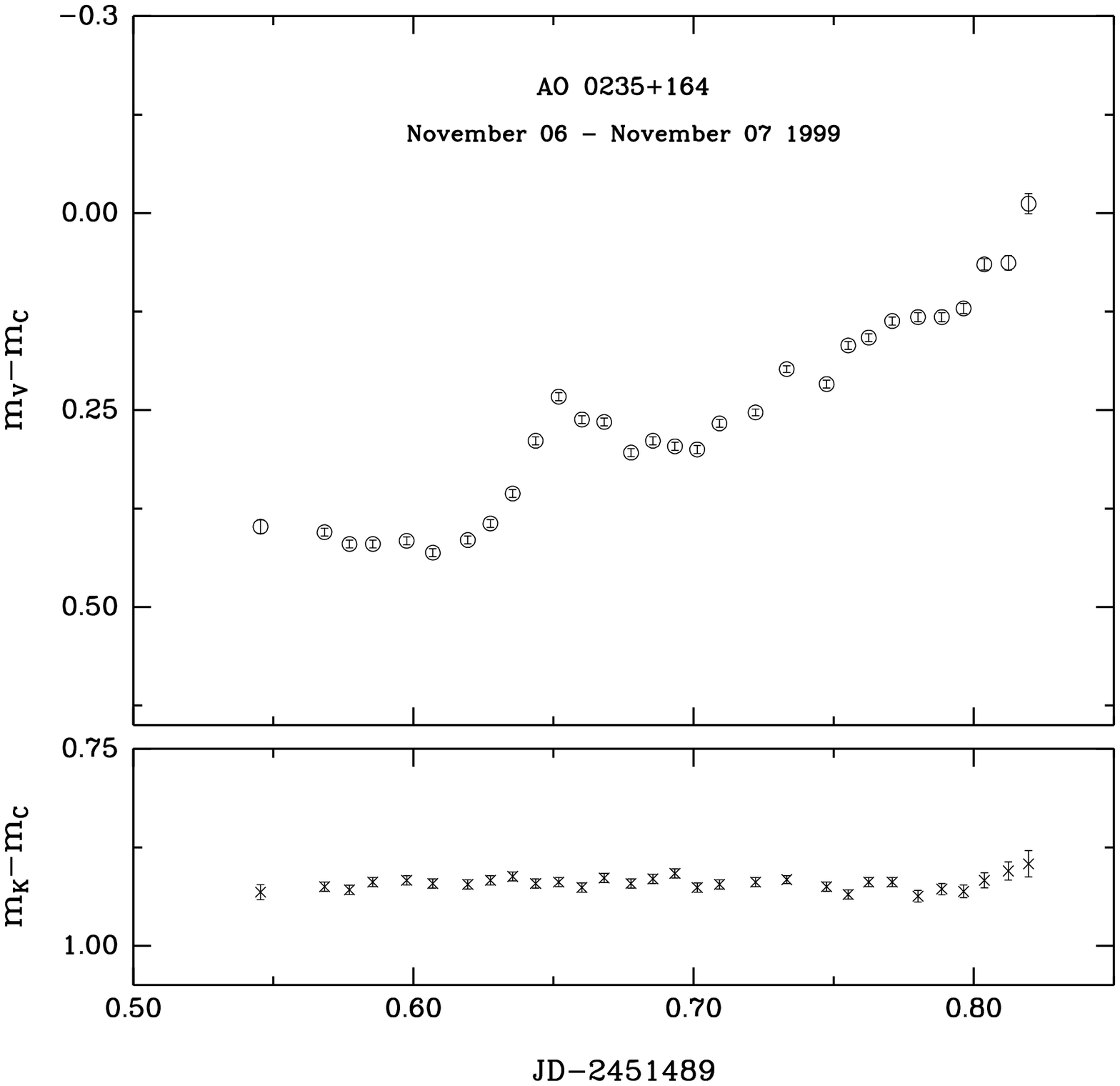}
\includegraphics[width=0.45\textwidth]{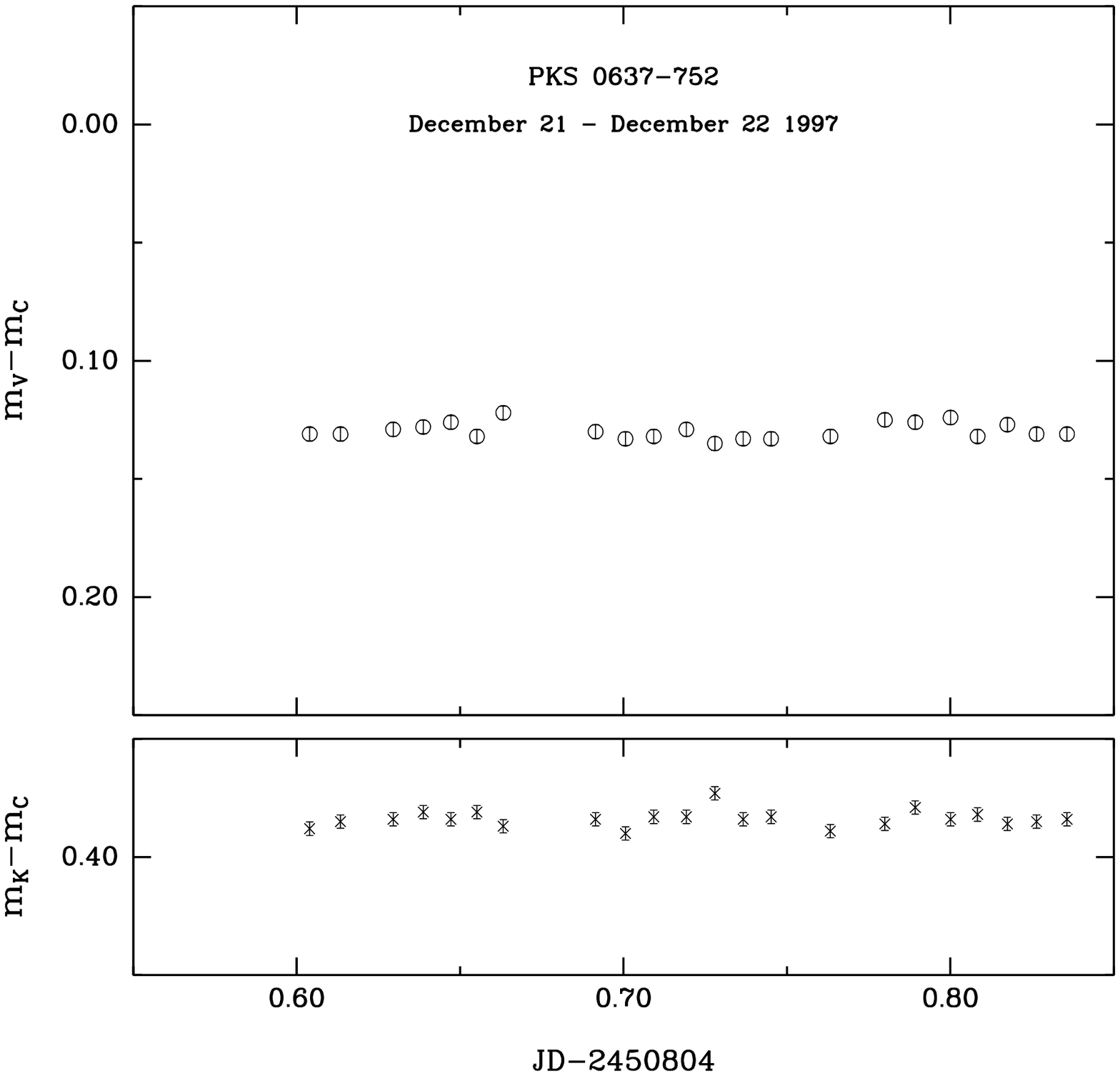}
\caption{Upper panel: DLCs for AO\,0235+164,
    showing strong variability. Lower panel: light curves for PKS
    0637$-$752, undetected variability. In both cases, we present $V$ filter
    observations, for $m_V - m_C$ (top) and $m_K - m_C$ (bottom).}
\label{fig:examp}
\end{figure}

In this work, unlike what was done by \citet{1999A&AS..135..477R}, who
constructed \lq mean\rq~comparison and control stars from three stars
in each frame, we followed the re\-com\-men\-da\-tion given by
\citet{1988AJ.....95..247H}, who used one comparison and one control
stars.  The criterion proposed by these authors suggests that the
magnitude of the control star must be as similar as possible to the
magnitude of the object, meanwhile for the comparison star, the
magnitude should be slightly brighter than the other two. Comparing
both criteria, we found that the criterion established by
\citet{1988AJ.....95..247H} is more conservative than the one proposed
by \citet{1999A&AS..135..477R} \citep[see][]{2011BAAA...54..325Z}.
The use of mean stars improves the signal-to-noise (S/N) relation of the \lq
control$-$comparison\rq~light curves and this may lead to an
overestimation of the AGN va\-ria\-bi\-li\-ty. Thus, choosing a pair of
candidates to control and comparison stars, we ge\-ne\-ra\-ted the
DLCs (\lq object$-$comparison\rq~and \lq
control$-$comparison\rq~) using a reduction package of {\sc iraf} ({\sc
  apphot}), and we analysed both curves, searching for a \lq
control$-$comparison\rq~light curve with the minimum possible
dispersion, while, at the same time, fulfilling the above-explained
conditions. In Fig.~\ref{fig:examp}, we show two extreme examples of
the light curves obtained (the light curves are as fig.~1 in
\citealp{2000A&A...360L..47R} and fig.~4 in
\citealp{1999A&AS..135..477R}, respectively).

\section{Statistical tests to study variability}
\label{sec:stsv}

In this section, we will analyse two
statistical methods most widely used to quantify variability in AGN 
light curves: the $C$ and $F$ statistics.

\subsection{{\boldmath $C$} criterion}
\label{ctest_def}

This is a criterion that contemplates the ratio of the standard
devia\-tions of the \lq object$-$comparison\rq~and \lq
control$-$comparison\rq~light curves, $\sigma_1$ and $\sigma_2$
respectively; the $C$ parameter is defined as:

\begin{equation}
C=\frac {\sigma_1}{\sigma_2}.
\label{eq:C}
\end{equation}

If $C$ is greater than a critical value (i.e. $C \ge 2.576$), the
light curve of the source is said to be variable with a 99.5 per cent
confidence level (CL).

\subsubsection{Scaled $C$ criterion}
\label{sec:sCc}

\citet{1988AJ.....95..247H} define a scale factor, $\Gamma$, to be
applied when no comparison and control stars, meeting the criterion
mentioned in Section 2.1, are found in the field. It takes into
account the diffe\-rent relative brightnesses between the AGN and the
comparison and control stars. This is so because the budget of
photometric errors includes flux-dependent terms, as well as terms
that are the same for all objects, irrespective of their magnitudes
(sky and read-out noise).

This factor is given by 
\citet{1988AJ.....95..247H}, 

\begin{eqnarray}
\Gamma^{2} &=& \frac
      {{\sigma_{1}^{2}}_{\mathrm{(INST)}}}{\sigma_{2}^{2}}\\ \nonumber &=& \left
      (\frac {f_K}{f_V} \right)^{2} \left\lfloor \frac
      {f_C^{2}(f_V+P)+f_V^{2}(f_C+P)}{f_K^{2}(f_C+P)+f_C^{2}(f_K+P)}
      \right\rfloor
\end{eqnarray}

\noindent where $f_V, f_K, f_C$ are the fluxes in adu for the object, control
and comparison stars, respectively; and $P$ takes into account the sky
photons and the read-out noise. The scale factor calculation is made
by an estimation of the ratio between ${\sigma_1^2}_{\mathrm{(INST)}}$
(variance of the \lq object$-$comparison\rq~curve predicted by the
CCD-based error equation and the median V and C measurements) and
$\sigma_2^2$, through the properties of the CCD used (i.e. gain and
read-out noise), as well as a proper weighting of the counts for each
object and for the sky \citep[see][for
  details]{1988AJ.....95..247H}. Then, the scaled $C$ parameter
results:

\begin{equation}
C= \frac {\sigma_1}{\Gamma \sigma_2}.
\end{equation}

This weight factor is important since, in many cases, the fields are not
very populated, limiting the choice of the comparison and control stars. In
those cases, there is an error term that is an in\-crea\-sing function of the
difference between the magnitudes of the objects. The use of the $\Gamma$
factor compensates for such differences.

\subsection{{\boldmath $F$}-test statistic}

In this statistic, it is assumed that errors in the curves are
distributed normally and their associated distributions need not have
the same degrees of freedom. The parameter $F$ is defined as:

\begin{equation}
F= \frac{\sigma^{2}_{1}}{\sigma^{2}_{2}}
\end{equation}

\noindent where $\sigma^{2}_{1}$ is the variance of the \lq
object$-$comparison\rq~light curve, and $\sigma^{2}_{2}$ that of the \lq
control$-$comparison\rq~curve.

The calculated $F$ values are compared with critical values
$F^{\alpha}_{n_{VC},n_{KC}}$, which have an associated significance
level, $\alpha$, and degrees of freedom of the different
distributions. The degrees of freedom can be described as the number
of scores that are free to vary, while $1-\alpha$ is the cumulative
probability of the distribution.  In our case, the degrees of freedom
are associated with the number of points in the \lq
object$-$comparison\rq~light curve, $n_{VC}$, and in the \lq
control$-$comparison\rq, $n_{KC}$, where $n_{VC} = n_{KC} = n$,
resulting in $n-1$ degrees of freedom.

Then, if the parameter $F \ge F^{\alpha}_{n_{VC},n_{CK}}$, the null
hypothesis of the test (i.e. statistical equality between the
variances when there is no significant difference between them) is rejected,
meaning that the curve is classified as variable.

\subsubsection{Scaled $F$-test statistic}

As for the $C-$criterion, there is also a scaled version of the
$F-$test; in fact, this was the expression originally proposed by
\citet{1988AJ.....95..247H}. Thus, the weighted parameter $F$ is:

\begin{equation}
F= \frac{\sigma^{2}_{1}}{\Gamma^{2} \sigma^{2}_{2}} \; .
\end{equation}

\citet{2011MNRAS.412.2717J} propose an alternative to the $\Gamma$
corrective factor: they scale the variance $\sigma^{2}_{2}$ by a
factor $\kappa$, which is defined as the ratio of the average square
errors of the individual points in the DLCs.  The
main difference between $\Gamma$ and $\kappa$ is that the first is
obtained from mean values of object fluxes and sky counts for each
light curve, while the second takes into account individual error bars
for each data point. Since the relevant input parameters are basically
the same in both cases, they should provide similar results.

\subsection{Results and analysis}
\label{sec:r+a}

We present in Table \ref{tab:resul} the results of applying the $C$
criterion and the $F$ test to the sample of AGN light curves. We show
the object name, date, the number of points in the light curve ($n$),
the values of $C$ without/with weight ($C$ and $C_{\Gamma}$), the
va\-lues of $F$ without/with weight ($F$ and $F_{\Gamma}$), the
dispersion of the \lq control-comparison\rq~light curve multiplied by
$\Gamma$ and the weight factor $\Gamma$. The last column gives the
area to the left of the observed $F$ below the $F$ density distribution,
for the adopted 99.5 per cent$-$CL. A value of area-$F_{\Gamma} > 0.995$
means that the null$-$hypothesis (non-variable) should be rejected.

\begin{table*}
\begin{minipage}{130mm}
\caption[Results of the $C$ criterion and the $F$ test]{Results of the
  $C$ criterion and the $F$ test.  The columns are object; date;
  number of points, $n$; values of $C$ without/with weight, $C$ and
  $C_{\Gamma}$; values of $F$ without/with weight, $F$ and
  $F_{\Gamma}$; the dispersion of the \lq control$-$comparison\rq~light
  curve multiplied by $\Gamma$, the weight factor, $\Gamma$ and the
  area to the left of the observed $F$ below the $F$ density
  distribution, area-$F_{\Gamma}$. Numbers in boldface indicate
  variability.}
\label{tab:resul}
\begin{tabular}{rccccccccc}
\noalign{\medskip} \hline \noalign{\smallskip}
Object & Date & $n$ & $C$ & $C_{\Gamma}$ & $F$ & $F_{\Gamma}$  & $\Gamma \sigma_2$ & $\Gamma$ & Area-$F_{\Gamma}$\\
\noalign{\smallskip} \hline \noalign{\smallskip}
0208$-$512   & 11/03/99 & 40 & {\bf \phantom{0}9.34} & {\bf\phantom{0}9.61} & {\bf\phantom{0}87.32}            & {\bf\phantom{0}92.34}           & 0.005 & 0.973 & 1.0000\\
             & 11/04/99 & 39 & \phantom{0}2.00       & \phantom{0}2.15      & {\bf\phantom{0}\phantom{0}4.02}  & {\bf\phantom{0}\phantom{0}4.60} & 0.003 & 0.934 & 1.0000\\
0235$+$164   & 11/03/99 & 23 & {\bf 10.10}           & {\bf 11.47}          & {\bf 102.00}                     & {\bf 131.60}                    & 0.013 & 0.880 & 1.0000\\
             & 11/04/99 & 22 & {\bf\phantom{0}6.10}  & {\bf\phantom{0}5.66} & {\bf\phantom{0}37.22 }           & {\bf\phantom{0}32.06}           & 0.130 & 1.078 & 1.0000\\
             & 11/05/99 & 27 & {\bf 12.32}           & {\bf12.66}           & {\bf 151.65}                     & {\bf 160.3}                     & 0.007 & 0.973 & 1.0000\\
             & 11/06/99 & 22 & {\bf\phantom{0}4.37}  & {\bf\phantom{0}2.93} & {\bf\phantom{0}19.10}            & {\bf\phantom{0}\phantom{0}8.60} & 0.010 & 1.492 & 1.0000\\
             & 11/07/99 & 30 & {\bf 14.34}           & {\bf 17.74}          & {\bf 205.60}                     & {\bf 314.62}                    & 0.007 & 0.808 & 1.0000\\
             & 11/08/99 & 12 & {\bf\phantom{0}2.75}  & {\bf\phantom{0}2.95} & {\bf\phantom{0}\phantom{0}7.56}  & {\bf\phantom{0}\phantom{0}8.70} & 0.009 & 0.933 & 0.9988\\
             & 12/22/00 & 10 & {\bf\phantom{0}3.30}  & {\bf\phantom{0}3.44} & {\bf\phantom{0}10.90}            & {\bf\phantom{0}11.83}           & 0.007 & 0.959 & 0.9989\\
             & 12/24/00 & 11 & {\bf\phantom{0}5.55}  & {\bf\phantom{0}6.65} & {\bf\phantom{0}30.81}            & {\bf\phantom{0}44.20}           & 0.008 & 0.835 & 1.0000\\
0521$-$365   & 12/17/98 & 29 & {\bf\phantom{0}3.90}  & {\bf\phantom{0}4.50} & {\bf\phantom{0}15.14}            & {\bf\phantom{0}20.27}           & 0.004 & 0.864 & 1.0000\\
0537$-$441   & 12/22/97 & 23 & {\bf\phantom{0}5.85}  & {\bf\phantom{0}4.67} & {\bf\phantom{0}34.25}            & {\bf\phantom{0}21.85}           & 0.005 & 1.252 & 1.0000\\
             & 12/23/97 & 23 & {\bf\phantom{0}4.30}  & {\bf\phantom{0}3.67} & {\bf\phantom{0}18.46}            & {\bf\phantom{0}13.47}           & 0.005 & 1.171 & 1.0000\\
             & 12/16/98 & 35 & {\bf\phantom{0}4.96}  & {\bf\phantom{0}5.93} & {\bf\phantom{0}24.63}            & {\bf\phantom{0}35.22}           & 0.004 & 0.836 & 1.0000\\
             & 12/17/98 & 33 & {\bf\phantom{0}6.28}  & {\bf\phantom{0}6.98} & {\bf\phantom{0}39.46}            & {\bf\phantom{0}48.82}           & 0.005 & 0.899 & 1.0000\\
             & 12/18/98 & 55 & \phantom{0}1.50       & \phantom{0}1.60      & {\bf\phantom{0}\phantom{0}2.24}  & {\bf\phantom{0}\phantom{0}2.57} & 0.004 & 0.932 & 0.9993\\
             & 12/19/98 & 14 & \phantom{0}1.77       & \phantom{0}1.98      & \phantom{0}\phantom{0}3.12       & \phantom{0}\phantom{0}3.93      & 0.011 & 0.891 & 0.9805\\
             & 12/21/98 & 42 & \phantom{0}1.92       & \phantom{0}2.31      & {\bf\phantom{0}\phantom{0}3.69}  & {\bf\phantom{0}\phantom{0}5.33} & 0.004 & 0.832 & 1.0000\\
             & 12/20/00 & 11 & \phantom{0}1.01       & \phantom{0}1.61      & \phantom{0}\phantom{0}1.01       & \phantom{0}\phantom{0}2.61      & 0.006 & 0.624 & 0.8534\\
             & 12/21/00 & 41 & \phantom{0}0.72       & \phantom{0}1.51      & \phantom{0}\phantom{0}1.91       & \phantom{0}\phantom{0}1.33      & 0.004 & 0.628 & 0.6245\\
             & 12/22/00 & 46 & \phantom{0}0.47       & \phantom{0}0.75      & {\bf\phantom{0}\phantom{0}4.54}  & \phantom{0}\phantom{0}1.80      & 0.006 & 0.630 & 0.9488\\
             & 12/23/00 & 57 & \phantom{0}0.97       & \phantom{0}1.54      & \phantom{0}\phantom{0}1.07       & {\bf\phantom{0}\phantom{0}2.37} & 0.004 & 0.629 & 0.9984\\
             & 12/24/00 & 50 & \phantom{0}1.12       & \phantom{0}1.79      & \phantom{0}\phantom{0}1.26       & {\bf\phantom{0}\phantom{0}3.21} & 0.004 & 0.627 & 0.9999\\
0637$-$752   & 12/21/97 & 22 & \phantom{0}0.95       & \phantom{0}0.93      & \phantom{0}\phantom{0}1.10       & \phantom{0}\phantom{0}1.15      & 0.004 & 1.021 & 0.2514\\
             & 12/22/97 & 26 & \phantom{0}0.97       & \phantom{0}0.95      & \phantom{0}\phantom{0}1.05       & \phantom{0}\phantom{0}1.10      & 0.004 & 1.023 & 0.1890\\
1034$-$293   & 04/24/97 & 15 & \phantom{0}1.97       & \phantom{0}1.86      & \phantom{0}\phantom{0}3.89       & \phantom{0}\phantom{0}3.46      & 0.014 & 1.060 & 0.9731\\
1101$-$232   & 04/29/98 & 32 & \phantom{0}0.73       & \phantom{0}0.74      & \phantom{0}\phantom{0}1.88       & \phantom{0}\phantom{0}1.81      & 0.006 & 0.979 & 0.8962\\
1120$-$272   & 04/27/98 & 15 & \phantom{0}0.62       & \phantom{0}0.67      & \phantom{0}\phantom{0}2.57       & \phantom{0}\phantom{0}2.24      & 0.054 & 0.934 & 0.8558\\
1125$-$305   & 04/28/97 & 35 & \phantom{0}0.96       & \phantom{0}0.97      & \phantom{0}\phantom{0}1.09       & \phantom{0}\phantom{0}1.06      & 0.009 & 0.987 & 0.1286\\
1127$-$145   & 04/27/98 & 14 & \phantom{0}1.31       & \phantom{0}1.23      & \phantom{0}\phantom{0}1.72       & \phantom{0}\phantom{0}1.51      & 0.004 & 1.068 & 0.5300\\
1144$-$379   & 04/27/97 & 39 & \phantom{0}1.84       & \phantom{0}1.21      &{\bf \phantom{0}\phantom{0}3.40}  & \phantom{0}\phantom{0}1.47      & 0.029 & 1.521 & 0.7573\\
1157$-$299   & 04/28/98 & 26 & \phantom{0}0.73       & \phantom{0}0.84      & \phantom{0}\phantom{0}1.86       & \phantom{0}\phantom{0}1.41      & 0.005 & 0.870 & 0.6006\\
1226$+$023   & 04/08/00 & 26 & \phantom{0}1.04       & \phantom{0}1.44      & \phantom{0}\phantom{0}1.09       & \phantom{0}\phantom{0}2.07      & 0.003 & 0.724 & 0.9266\\
             & 04/09/00 & 22 & \phantom{0}1.02       & \phantom{0}1.41      & \phantom{0}\phantom{0}1.04       & \phantom{0}\phantom{0}2.00      & 0.004 & 0.720 & 0.8793\\
1229$-$021   & 04/11/00 & 24 & \phantom{0}1.27       & \phantom{0}1.32      & \phantom{0}\phantom{0}1.62       & \phantom{0}\phantom{0}1.74      & 0.007 & 0.965 & 0.8095\\
             & 04/12/00 & 25 & \phantom{0}1.82       & \phantom{0}1.87      & {\bf\phantom{0}\phantom{0}3.32}  & {\bf\phantom{0}\phantom{0}3.51} & 0.005 & 0.972 & 0.9969\\
1243$-$072   & 04/08/00 & 24 & \phantom{0}1.48       & \phantom{0}0.97      & \phantom{0}\phantom{0}2.19       & \phantom{0}\phantom{0}1.06      & 0.038 & 1.523 & 0.1098\\
             & 04/09/00 & 24 & \phantom{0}2.24       & \phantom{0}1.45      & {\bf\phantom{0}\phantom{0}5.03}  & \phantom{0}\phantom{0}2.11      & 0.032 & 1.542 & 0.9209\\
1244$-$255   & 04/29/98 & 26 & {\bf\phantom{0}4.40}  & {\bf\phantom{0}4.53} & {\bf\phantom{0}19.30}            & {\bf\phantom{0}20.51}           & 0.005 & 0.970 & 1.0000\\
1253$-$055   & 06/08/99 & 22 & \phantom{0}1.16       & \phantom{0}1.57      & \phantom{0}\phantom{0}1.35       & \phantom{0}\phantom{0}2.45      & 0.011 & 0.743 & 0.9544\\
1256$-$229   & 04/24/98 & 20 & \phantom{0}1.49       & \phantom{0}1.74      & \phantom{0}\phantom{0}2.21       & \phantom{0}\phantom{0}3.05      & 0.005 & 0.852 & 0.9806\\
1331$+$170   & 04/10/00 & 30 & \phantom{0}1.17       & \phantom{0}1.17      & \phantom{0}\phantom{0}1.40       & \phantom{0}\phantom{0}1.36      & 0.007 & 1.003 & 0.5924\\
1334$-$127   & 04/11/00 & 30 & {\bf\phantom{0}2.87}  & {\bf\phantom{0}3.72} & {\bf\phantom{0}\phantom{0}8.23}  & {\bf\phantom{0}13.87}           & 0.005 & 0.770 & 1.0000\\
             & 04/12/00 & 31 & \phantom{0}2.42       & {\bf\phantom{0}2.97} & {\bf\phantom{0}\phantom{0}5.85}  & {\bf\phantom{0}\phantom{0}8.81} & 0.008 & 0.815 & 1.0000\\
1349$-$439   & 04/24/98 & 14 & \phantom{0}2.11       & \phantom{0}2.16      & \phantom{0}\phantom{0}4.46       & \phantom{0}\phantom{0}4.66      & 0.009 & 0.979 & 0.9908\\
1424$-$418   & 06/04/99 & 15 & \phantom{0}1.56       & \phantom{0}1.78      & \phantom{0}\phantom{0}2.42       & \phantom{0\phantom{0}}3.17      & 0.021 & 0.874 & 0.9614\\
             & 06/05/99 & 19 & \phantom{0}0.74       & \phantom{0}0.81      & \phantom{0}\phantom{0}1.84       & \phantom{0}\phantom{0}1.53      & 0.032 & 0.911 & 0.6224\\
1510$-$089   & 04/29/98 & 25 & \phantom{0}1.13       & \phantom{0}1.17      & \phantom{0}\phantom{0}1.28       & \phantom{0}\phantom{0}1.38      & 0.005 & 0.965 & 0.5596\\
             & 04/30/98 & 21 & \phantom{0}1.03       & \phantom{0}1.08      & \phantom{0}\phantom{0}1.06       & \phantom{0}\phantom{0}1.16      & 0.009 & 0.956 & 0.2537\\
             & 06/06/99 & 17 & \phantom{0}1.20       & \phantom{0}1.75      & \phantom{0}\phantom{0}1.45       & \phantom{0}\phantom{0}3.07      & 0.005 & 0.688 & 0.9687\\
             & 06/07/99 & 27 & \phantom{0}0.94       & \phantom{0}1.40      & \phantom{0}\phantom{0}1.14       & \phantom{0}\phantom{0}1.93      & 0.007 & 0.674 & 0.9015\\
1606$+$106   & 07/23/01 & 10 & \phantom{0}1.19       & \phantom{0}1.00      & \phantom{0}\phantom{0}1.42       & \phantom{0}\phantom{0}1.01      & 0.010 & 1.950 & 0.0076\\
             & 07/24/01 & 9  & \phantom{0}1.39       & \phantom{0}1.20      & \phantom{0}\phantom{0}1.92       & \phantom{0}\phantom{0}1.43      & 0.016 & 1.158 & 0.3783\\
1622$-$297   & 06/04/99 & 13 & {\bf 11.61}           & {\bf 11.50}          & {\bf 134.90}                     & {\bf 132.3}                     & 0.025 & 1.010 & 1.0000\\
             & 06/05/99 & 22 & \phantom{0}2.25       & \phantom{0}2.24      & {\bf\phantom{0}\phantom{0}5.07}  & {\bf\phantom{0}\phantom{0}5.01} & 0.015 & 1.006 & 0.9995\\
1741$-$038   & 06/06/99 & 20 & \phantom{0}1.57       & \phantom{0}1.31      & \phantom{0}\phantom{0}2.52       & \phantom{0}\phantom{0}1.73      & 0.024 & 1.206 & 0.7579\\
             & 06/07/99 & 22 & \phantom{0}2.20       & \phantom{0}1.76      & {\bf\phantom{0}\phantom{0}4.84}  & \phantom{0}\phantom{0}3.11      & 0.034 & 1.248 & 0.9877\\
1933$-$400   & 07/23/01 & 20 & \phantom{0}1.31       & \phantom{0}1.28      & \phantom{0}\phantom{0}1.73       & \phantom{0}\phantom{0}1.64      & 0.010 & 1.027 & 0.7098\\
             & 07/24/01 & 20 & \phantom{0}1.01       & \phantom{0}0.99      & \phantom{0}\phantom{0}1.03       & \phantom{0}\phantom{0}1.01      & 0.016 & 1.019 & 0.0158\\
\noalign{\smallskip}\hline
\end{tabular}
\end{minipage}
\end{table*}

\setcounter{table}{1}

\begin{table*}
\begin{minipage}{100mm}
\caption{Results of the $C$ criterion and the $F$ test. (\it Cont.)}
\begin{tabular}{rccccccccc}
\noalign{\medskip} \hline \noalign{\smallskip}
Object & Date & $n$ & $C$ & $C_{\Gamma}$ & $F$ & $F_{\Gamma}$ & $\Gamma \sigma_2$ & $\Gamma$ & Area-$F_{\Gamma}$\\
\noalign{\smallskip} \hline \noalign{\smallskip}
2005$-$489  & 04/26/97 & 45 & \phantom{0}1.12        & \phantom{0}1.60      & \phantom{0}\phantom{0}1.24       & {\bf\phantom{0}\phantom{0}2.56} & 0.003 & 0.697 & 0.9977\\
2022$-$077  & 07/25/01 & 20 & {\bf\phantom{0}4.18}   & {\bf\phantom{0}4.13} & {\bf\phantom{0}17.45}            & {\bf\phantom{0}17.02}           & 0.010 & 1.013 & 1.0000\\
            & 07/26/01 & 19 & \phantom{0}2.27        & {\bf\phantom{0}2.78} & {\bf\phantom{0}\phantom{0}5.15}  & {\bf\phantom{0}\phantom{0}7.71} & 0.010 & 0.817 & 0.9999\\
2155$-$304  & 07/27/97 & 74 & \phantom{0}0.95        & \phantom{0}1.82      & \phantom{0}\phantom{0}1.11       & {\bf\phantom{0}\phantom{0}3.31} & 0.007 & 0.521 & 1.0000\\
2200$-$181  & 07/26/97 & 33 & \phantom{0}1.17        & \phantom{0}1.54      & \phantom{0}\phantom{0}1.37       & \phantom{0}\phantom{0}2.37      & 0.003 & 0.761 & 0.9828\\
            & 07/27/97 & 37 & \phantom{0}0.87        & \phantom{0}1.16      & \phantom{0}\phantom{0}1.31       & \phantom{0}\phantom{0}1.34      & 0.002 & 0.757 & 0.6110\\
2230$+$114  & 07/23/01 & 18 & \phantom{0}1.76        & \phantom{0}1.17      & \phantom{0}\phantom{0}3.09       & \phantom{0}\phantom{0}1.36      & 0.008 & 1.505 & 0.4691\\
            & 07/24/01 & 18 & {\bf 11.06}            & {\bf\phantom{0}8.04} & {\bf 122.30}                     & {\bf\phantom{0}64.63}           & 0.006 & 1.376 & 1.0000\\
            & 07/25/01 & 8  & {\bf\phantom{0}7.10}   & {\bf\phantom{0}6.80} & {\bf\phantom{0}50.46}            & {\bf\phantom{0}46.10}           & 0.006 & 1.046 & 1.0000\\
2254$-$204  & 09/20/97 & 35 & \phantom{0}0.75        & \phantom{0}0.94      & \phantom{0}\phantom{0}1.80       & \phantom{0}\phantom{0}1.13      & 0.021 & 0.794 & 0.2850\\
2316$-$423  & 09/04/97 & 37 & \phantom{0}1.31        & \phantom{0}1.52      & \phantom{0}\phantom{0}1.72       & \phantom{0}\phantom{0}2.30      & 0.018 & 0.864 & 0.9653\\
            & 09/05/97 & 36 & \phantom{0}1.32        & \phantom{0}1.50      & \phantom{0}\phantom{0}1.75       & \phantom{0}\phantom{0}2.25      & 0.015 & 0.883 & 0.9827\\
2320$-$035  & 07/25/01 & 17 & \phantom{0}1.55        & \phantom{0}1.50      & \phantom{0}\phantom{0}2.41       & \phantom{0}\phantom{0}2.24      & 0.005 & 1.038 & 0.8729\\
            & 07/26/01 & 7  & \phantom{0}2.44        & \phantom{0}2.37      & \phantom{0}\phantom{0}5.96       & \phantom{0}\phantom{0}5.60      & 0.004 & 1.032 & 0.9452\\
2340$-$469  & 09/04/97 & 36 & \phantom{0}1.69        & \phantom{0}1.64      & {\bf\phantom{0}\phantom{0}2.85}  & {\bf\phantom{0}\phantom{0}2.70} & 0.007 & 1.026 & 0.9958\\
            & 09/05/97 & 38 & \phantom{0}0.94        & \phantom{0}0.92      & \phantom{0}\phantom{0}1.13       & \phantom{0}\phantom{0}1.19      & 0.008 & 1.027 & 0.3978\\
2341$-$444  & 09/17/97 & 48 & \phantom{0}0.92        & \phantom{0}0.92      & \phantom{0}\phantom{0}1.17       & \phantom{0}\phantom{0}1.18      & 0.023 & 1.003 & 0.4235\\
2344$-$465  & 09/19/97 & 53 & \phantom{0}0.99        & \phantom{0}0.95      & \phantom{0}\phantom{0}1.00       & \phantom{0}\phantom{0}1.01      & 0.010 & 1.044 & 0.2572\\
2347$-$437  & 09/18/97 & 56 & \phantom{0}1.05        & \phantom{0}0.99      & \phantom{0}\phantom{0}1.11       & \phantom{0}\phantom{0}1.02      & 0.009 & 1.068 & 0.0738\\
\noalign{\smallskip}\hline                                                                                                                 
\end{tabular}                                                                          \end{minipage}
\end{table*}

To compare the results of both tests, we considered the $C$ criterion
and $F$ test both without the weight factor and with weighted
statistics.  We found that considering the non-weighted statistics,
among the 25.64 per cent of the DLCs classified as variable applying
the $C$ parameter, all of them maintained the classification with the
$F$ test; while for the remaining 74.36 per cent of the DLCs
classified as non-variable with $C$, 20.68 per cent of them changed
its classification using the $F$ test. Regarding the weighted
statistics, within the 28.21 per cent of the DLCs classified as
variable with the $C$ criterion, again all of them maintained the
classification with the $F$ test; meanwhile, within the 71.79 per cent
of the DLCs classified as non-variable with the $C$ criterion, 19.54
per cent of them have been classified in the same way using the $F$
test. We want to note that the direction of change in the
classification is in one way: from {\it non-variable} with the $C$
criterion to {\it variable} with the $F$ test.  So, a significant
fraction of the curves that are classified as non-variable applying
the $C$ criterion, are classified as variable with the $F$ test, which
could indicate a higher sensitivity of the $F$ test (or, conversely, a
more conservative behaviour of the $C$ criterion).

\begin{figure}
\centering
\includegraphics[width=0.5\textwidth]{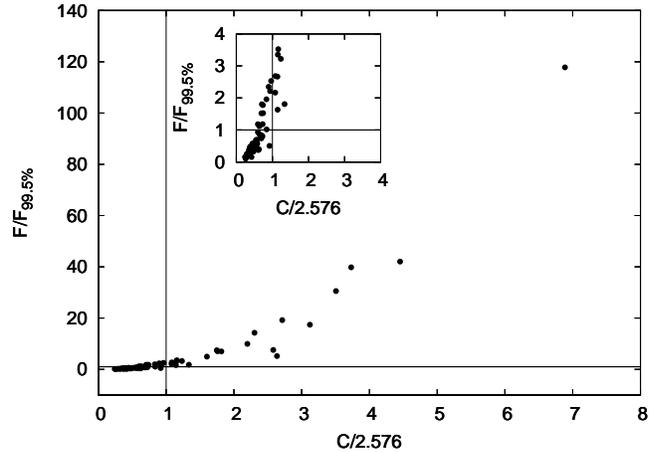}
\caption{Comparison between the $C$ and $F$ statistics (99.5 per cent
  sig\-ni\-fi\-can\-ce level).  A zoom of the region close to (1,1) is shown
  as an inset. Solid lines indicate the threshold of the critical
  values for both statistics.}
\label{fig:corrCF}
\end{figure}

Besides the adopted CL, we studied the behaviour of both statistics
relaxing the CL: 99.0 per cent and 95.0 per cent (the meaning of CL
for the $C$ criterion will be explained in Section~\ref{sec:iicc}).
As an example, in Fig. \ref{fig:corrCF} we present a comparison
between the values obtained for the weighted $C$ and $F$ parameters at
99.5 per cent of CL. These values were referred to the corres\-ponding
limiting values in each particular case in order to better compare
each other. Solid lines indicate the threshold of the critical values
for both statistics, marking the division for the four possible
cases. It is possible to appreciate that the quarter, in which the $C$
criterion would result variable and the $F$ test would not, is empty,
in contrast with the opposite quarter (non-variable with $C$, and
variable with $F$).

\subsection{Distributions}
\label{sec:distri}

\begin{figure}
\includegraphics[width=80mm]{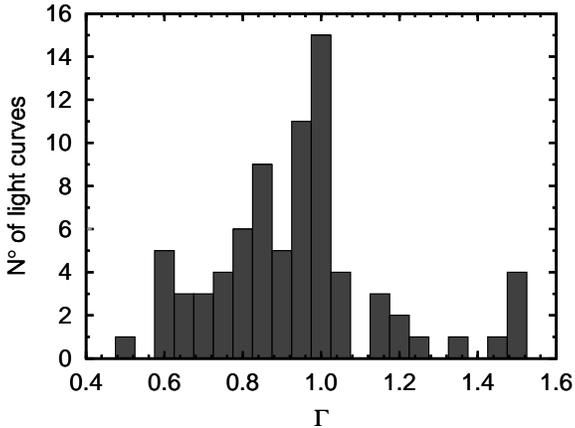}
\caption{Histogram of the values of $\Gamma$.}
\label{fig:Gamma}
\end{figure}

As we mentioned in Section~\ref{sec:sCc}, a scale factor was
introduced in order to compensate the differences in magnitude due to
the non$-$optimal choice of the comparison and control stars. In
Fig.~\ref{fig:Gamma}, we present the distribution of values of the
weight factor, $\Gamma$, obtained for each DLC.  It shows that the
peak in the distribution falls at $\Gamma=1$ and, taking an interval
of $\pm 0.2$, almost a 75 per cent of the DLCs are within this
interval. Recalling its definition, values close to 1 indicate that
both stars meet fairly well the criterion proposed by
\citet{1988AJ.....95..247H}.  Thus, in our case, the selection of the
pair of stars was almost optimal for the majority of the DLCs.

To understand the above$-$described behaviour and to determine what
parameters make a light curve more susceptible to changes in its
variability classification, we analysed the distributions of the
number of DLCs against their amplitudes, $\Delta m$; the elapsed time
corresponding to $\Delta m$, $\Delta t$; the number of observations
made du\-ring the night (i.e. number of points in the curve), $n$; and
the dispersion in the \lq control$-$comparison\rq~light curve,
$\sigma_2$. From here on, we define \lq Var\rq~for variable and \lq
NVar\rq~for non-variable. We built the corresponding histograms for
three groups of DLCs: those two that maintained their classifications
using both tests (i.e. Var$\rightarrow$Var and NVar$\rightarrow$NVar),
and the third one that changed its classification (i.e. NVar for the
$C$ criterion $\rightarrow$ Var for the $F$ test). We do not find any
case corresponding to the change Var$\rightarrow$NVar. Also, we
considered the same cases without/with the scale factor $\Gamma$.

\begin{figure}
\includegraphics[width=70mm]{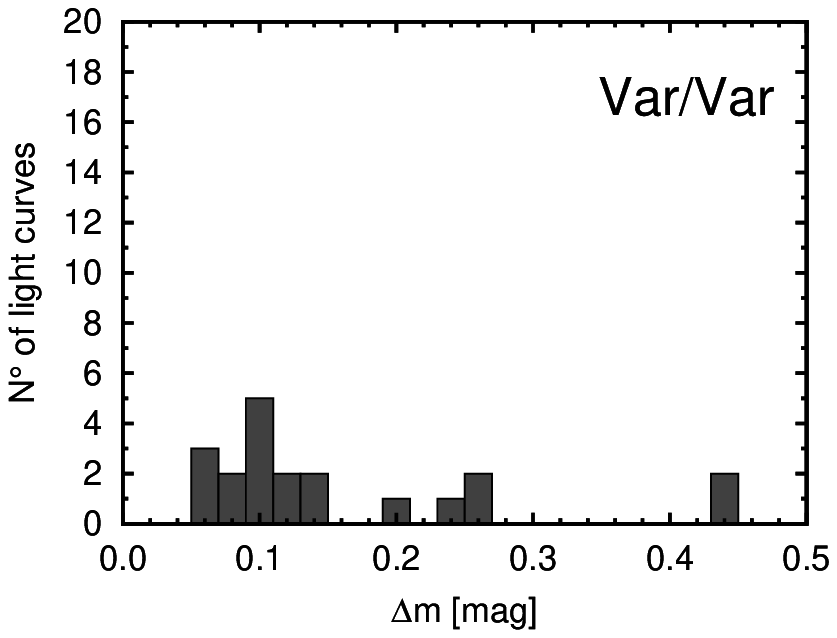}
\includegraphics[width=70mm]{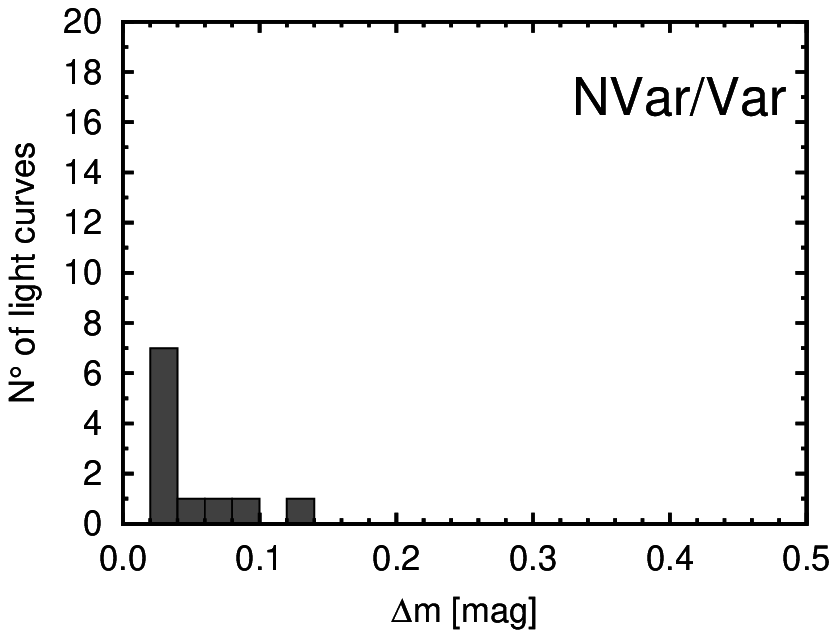}
\centering
\includegraphics[width=70mm]{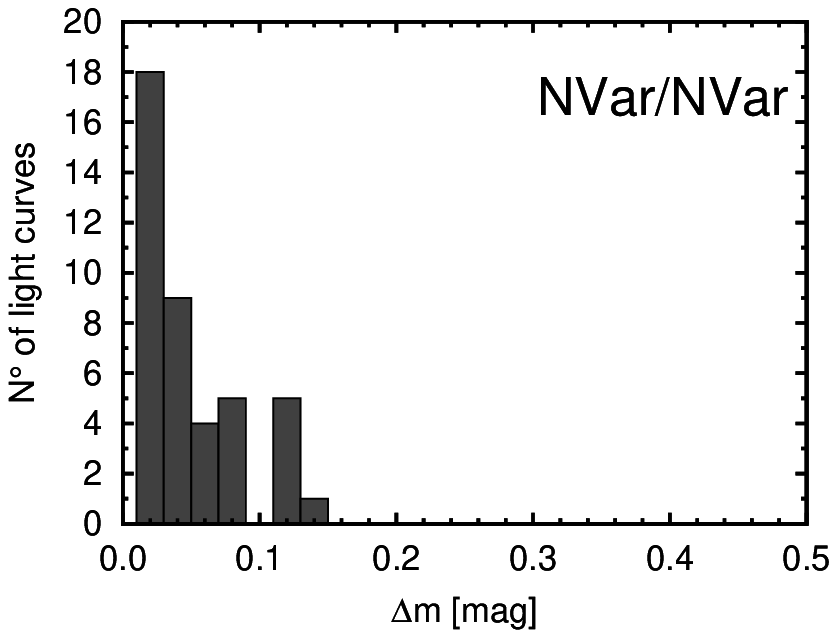}
\caption{Histograms of $\Delta m$ for the cases: Var/Var, NVar/Var and NVar/NVar.}
\label{fig:hDm}
\end{figure}

\begin{figure}
\includegraphics[width=70mm]{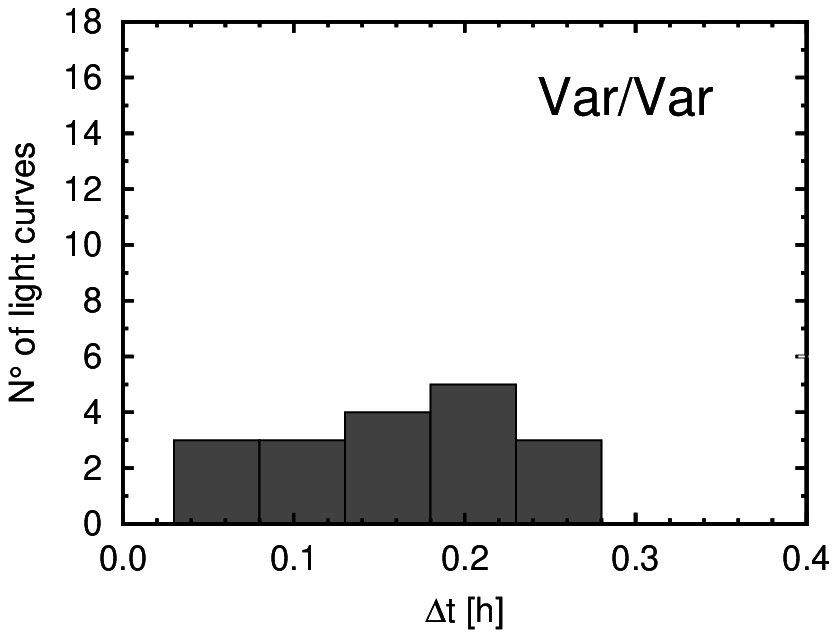}
\includegraphics[width=70mm]{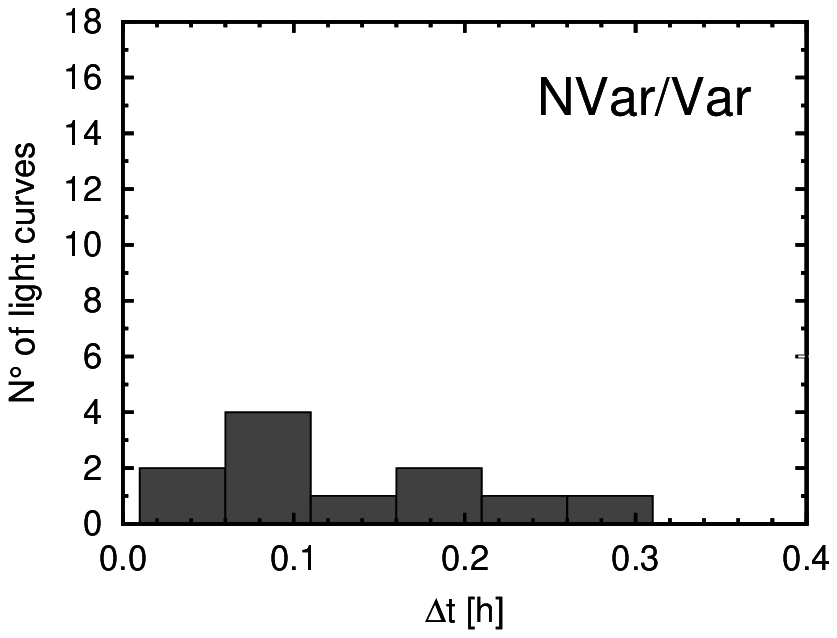}
\centering
\includegraphics[width=70mm]{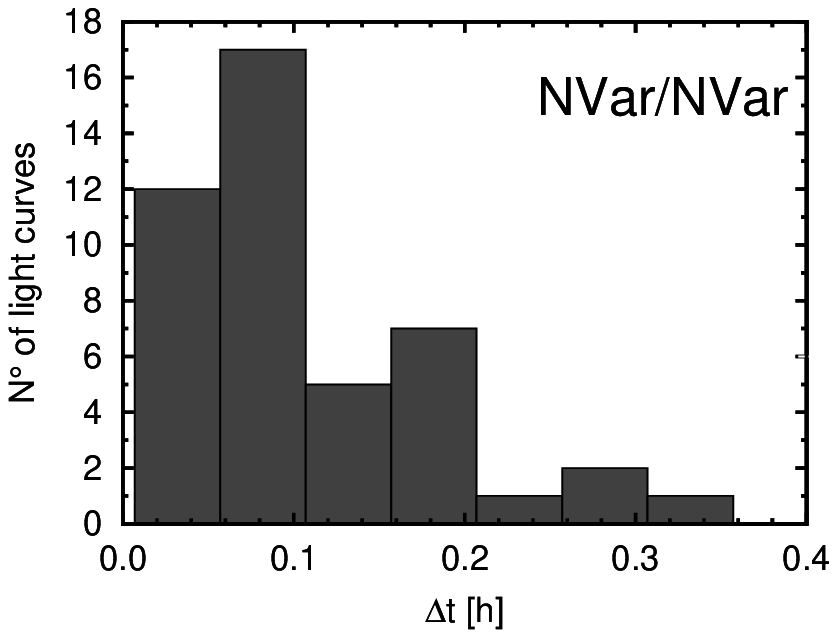}
\caption{Histograms of $\Delta t$ for the cases: Var/Var, NVar/Var and NVar/NVar.}
\label{fig:hDt}
\end{figure}

\begin{figure}
\includegraphics[width=70mm]{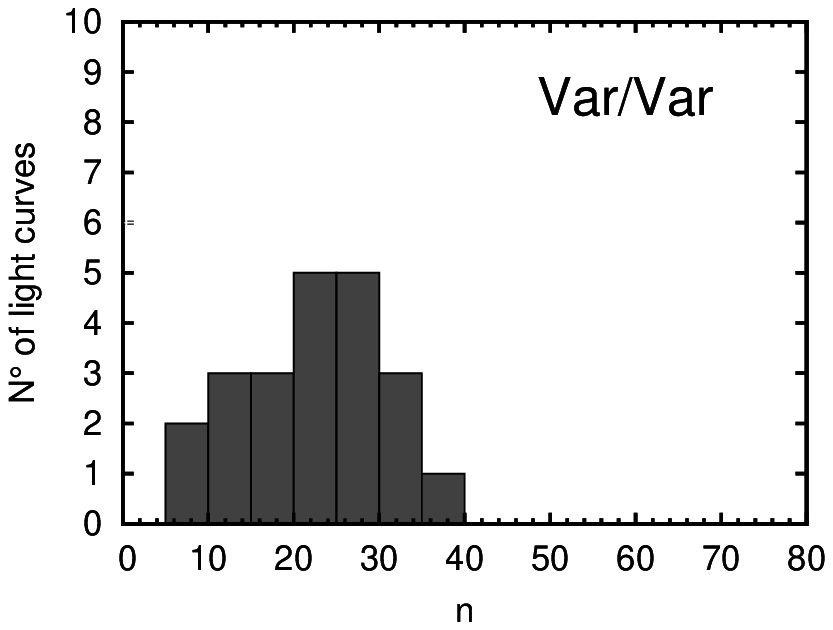}
\includegraphics[width=70mm]{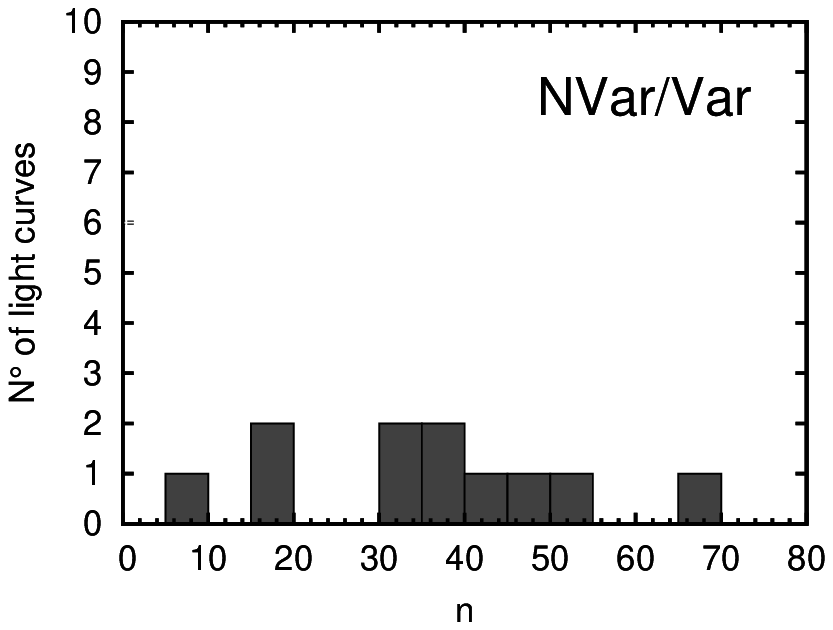}
\centering
\includegraphics[width=70mm]{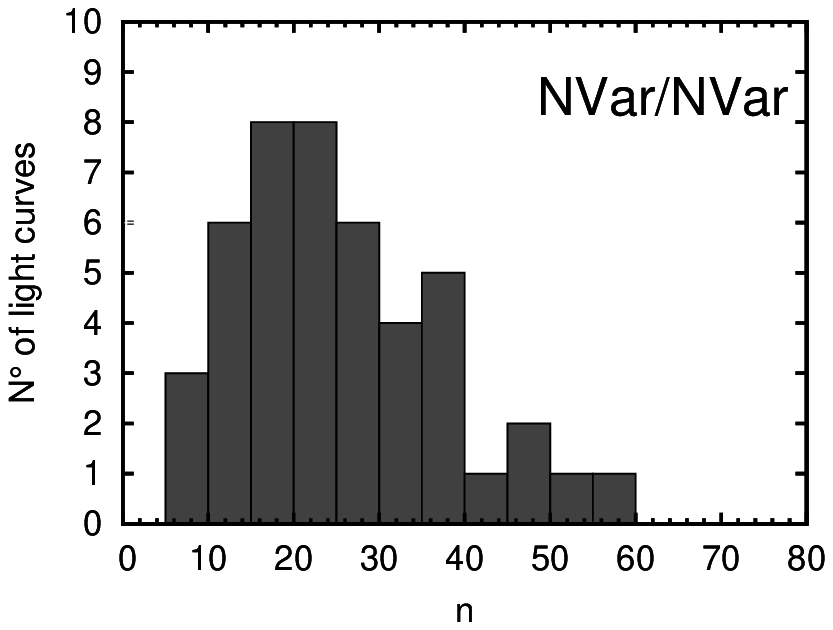}
\caption{Histograms of $n$ for the cases: Var/Var, NVar/Var and NVar/NVar.}
\label{fig:hn}
\end{figure}

\begin{figure}
\includegraphics[width=70mm]{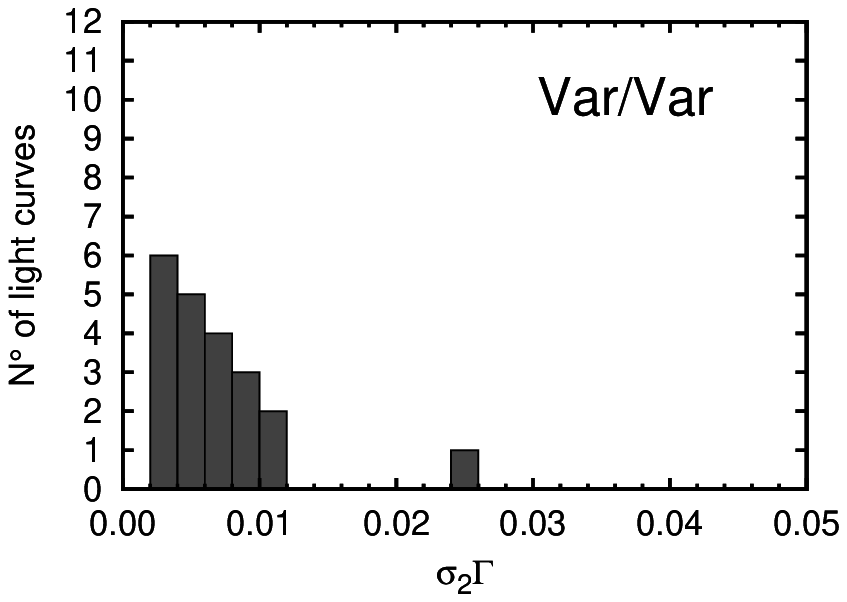}
\includegraphics[width=70mm]{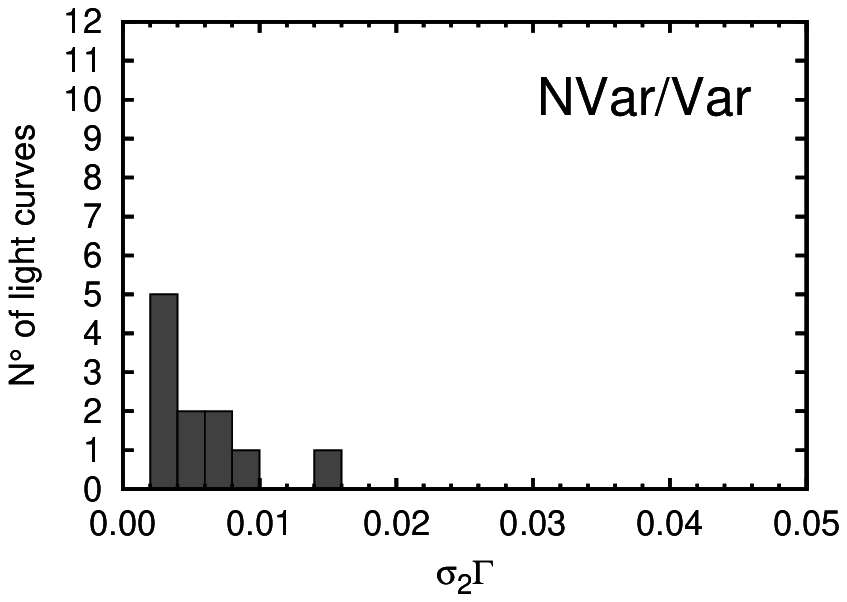}
\centering
\includegraphics[width=70mm]{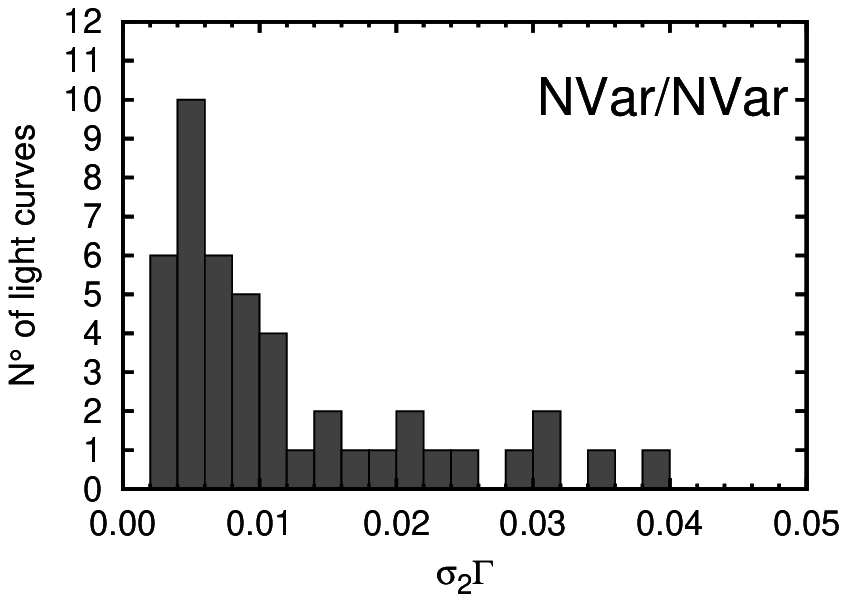}
\caption{Histograms of weighted $\sigma_2$ (i.e. $\sigma_2 \Gamma$)
  for the cases: Var/Var, NVar/Var and NVar/NVar.}
\label{fig:hs2}
\end{figure}

There is no significant difference between the distributions
without/with the factor $\Gamma$ (this is consistent with the fact
that $\langle \Gamma \rangle = 1$ with a small dispersion), so we
present only results inclu\-ding this factor. Note that this holds for
our particular DLC sample, for which $\langle \Gamma \rangle \approx
1$, but it will not be the case if control$-$comparison stars are not
suitably selected (i.e. $\langle \Gamma \rangle \gg 1$).  The
histograms presented in Fig.~\ref{fig:hDm} correspond to $\Delta m$,
to $\Delta t$ in Fig.~\ref{fig:hDt}, to $n$ in Fig.~\ref{fig:hn} and
to $\sigma_{2}$ in Fig.~\ref{fig:hs2}.

\begin{table*}
\begin{minipage}{100mm}
\caption{Results of the KS test. The columns show the
  variable considered; which distributions are compared; the KS
  statistical parameter $Z$; the maximum distance between
  distributions, $d$; and the area under the distribution of the
  statistical parameter $Z$ to the left, 1-prob.}
\label{tab:KS}
\begin{tabular}{ccccc}
\noalign{\medskip} \hline \noalign{\smallskip}
Variable & Compared distributions & $Z$ & $d$ & 1-prob \\
\noalign{\smallskip} \hline \noalign{\smallskip}
$\Delta m$     & Var/Var versus NVar/Var   & 2.0409 & 0.727 & 0.999\\
               & Var/Var versus NVar/NVar  & 2.5058 & 0.644 & 0.999\\
               & NVar/Var versus NVar/NVar & 0.6632 & 0.222 & 0.282\\
\noalign{\smallskip} \hline \noalign{\smallskip}              
$\Delta t$     & Var/Var versus NVar/Var   & 0.6373 & 0.227 & 0.211\\
               & Var/Var versus NVar/NVar  & 1.6226 & 0.417 & 0.992\\
               & NVar/Var versus NVar/NVar & 1.3084 & 0.438 & 0.954\\
\noalign{\smallskip} \hline \noalign{\smallskip}               
$n$            & Var/Var versus NVar/Var   & 1.9146 & 0.682 & 0.999\\
               & Var/Var versus NVar/NVar  & 0.7704 & 0.198 & 0.447\\
               & NVar/Var versus NVar/NVar & 1.5086 & 0.505 & 0.986\\
\noalign{\smallskip} \hline \noalign{\smallskip}                
$\sigma_{2}$   & Var/Var versus NVar/Var   & 1.2773 & 0.455 & 0.933\\
               & Var/Var versus NVar/NVar  & 1.0350 & 0.266 & 0.790\\
               & NVar/Var versus NVar/NVar & 1.2367 & 0.414 & 0.931\\
\noalign{\smallskip}\hline
\end{tabular}
\end{minipage}
\end{table*}

\subsection{Details on the distributions}
\label{sec:histos}

In order to statistically study the behaviour observed in the
histograms, we applied a {\it goodness-of-fit Kolmogorov$-$Smirnov test}
(KS) to the data used to build the histograms. The results are
presented in Table \ref{tab:KS}.  The columns show the variable
considered; the distributions compared; the KS statistical parameter
$Z$; the maximum distance between distributions, $d$; and the area
under the distribution of $Z$ to the left, 1-prob.

In the following, we analyse the results shown in
Figs.~\ref{fig:hDm}$-$\ref{fig:hs2}, and quantified in
Table~\ref{tab:KS}.

\noindent \textit{DLC amplitude:} the DLCs classified as
non-variable with both tests (NVar/NVar), as well as those that change
status depending on the criterion used (NVar/Var), show distributions
strongly concentrated to small $\Delta m$ values (Fig.~\ref{fig:hDm}).
The KS test gives a level of significance 1-prob$=0.282$; thus, it
cannot be said that both distributions are statistically different.
Both have a high peak at $\Delta m \approx 0.03$\,mag, a value near
the typical instrumental noise in light curves.  Several of these
light curves are identified as variable by the $F$ test, while none of
them passes the $C$ criterion (see the Var/Var panel in
Fig.~\ref{fig:hDm}).

\noindent DLCs with high $\Delta m$ values will thus tend to be
classified as varia\-ble with both parameters, while the $F$ test, in
particular, seems prone to classify as variable some DLCs with
amplitudes very near to the rms error.

\noindent \textit{Elapsed time:} DLCs classified as non-variable with
both parameters have a broad distribution, with a peak around low
values ($\Delta t$ $\le 0.1$ h; Fig.~\ref{fig:hDt}). This peak is
consistent with variations due to relatively rapid fluctuations of
atmospheric conditions and photometric errors.

\noindent Regarding the distributions of DLCs classified as variable
with the $F$ test (NVar/Var and Var/Var), they are wider, differing
significantly from the NVar/NVar case.  This agrees with the fact that
a high value of $\Delta t$ tends to be more characteristic of curves
that present a systematic variability as opposed to fast
instrumental/atmospheric flickering.  In those curves, where the
instrumental noise is relatively low, this fact is more noticeable.
While the $F$ test seems to be more sensitive to classify as variable
curves with these characteristics, the KS test gives 1-prob$=0.211$
for the Var/Var versus NVar/Var histograms (Figs~\ref{fig:hDt}a and b),
meaning that we cannot claim that the distributions are statistically
different.

\noindent\textit{Number of observations:} in the cases where the
classification does not change (Var/Var and NVar/NVar,
Figs~\ref{fig:hn}a and c), the distributions are broad, peaking at $n
\approx 20$, i.e. about the median number of data points in our
DLCs. The KS test gives 1-prob$=0.447$ for the Var/Var versus NVar/NVar
histograms. The NVar/Var case, in turn, shows a much flatter
distribution, indicating some preference in favour of heavily sampled
DLCs. This is usually the case of bright objects, for which exposure
times are short (a few minutes), and photometric errors are usually
smaller.

\noindent\textit{Dispersion of the control$-$comparison DLC:} in those
cases in which the state of variability is maintained (i.e., Var/Var
and NVar/NVar; Figs~\ref{fig:hs2}a and c), we observe that the
distributions of $\Gamma \sigma_{2}$ clump below $\sim 0.012$ mag.
This implies DLCs with low instrumental dispersion, i.e. with high S/N
ratio. The variability detection in these DLCs (non$-$detection in the
case of NVar/NVar) is thus robust. However, for the NVar/NVar case,
there is a tail of DLCs with $\Gamma \sigma_{2} \ge 0.02$\,mag. This
means low S/N ratio; hence, any intrinsic AGN variability of low
amplitude would be masked by the, relatively, high noise.

\noindent The distribution of NVar/NVar cases is broader than that for
Var/Var (the KS test gives a value 1-prob$= 0.790$, i.e. it cannot be
said that the Var/Var and NVar/NVar histograms are statistically
different). This would imply a slightly larger sensitivity of the $F$
test to detect variability in noisy DLCs (or, from a different point
of view, a higher tendency to produce false positives under low S/N
conditions).
  
We also made an analysis of the light curves obtained after
interchanging the roles of the comparison and control stars, in order
to study how the choice of these stars could influence the statistical
results.  We applied both parameters to the DLCs,
finding out that close to the 95 per cent of the light curves
maintained their classifications with the $C$ criterion; meanwhile, for
the $F$ test that percentage dropped to 85 per cent.  This is
consistent with the fact that the mean value of $\Gamma$ is close to
$1$, with a low dispersion.  However, again, $F$ seems more sensitive
to systematics than $C$.

\section{Inquiring into the $C$ criterion}
\label{sec:iicc}

As defined in Section \ref{ctest_def}, the parameter $C$ is the ratio
between the standard deviations of two given distributions.  The
genesis of its use in AGN microvariability studies can be traced back
to \citet{1990AJ....100..347C} who proposed that the dispersion of the
differential magnitudes of the control light curve could provide an
estimator for the stability of the standard stars used in the data
analysis, being a more reliable measure of the observational
uncertainty than formal photometric errors.  A further step was given
by \citet{1995ApJ...452..582J}; they fitted both \lq
object$-$comparison\rq~and \lq control$-$comparison\rq~light curves with
straight lines and computed the standard deviations of the data points
in each curve. The largest value, either from one or from the other
light curve, was taken as a measure of the observational error. Note
that this procedure removes any long-term variation in the light
curves, while, at the same time, is insensitive to any \lq erratic,
low$-$amplitude variation\rq~of the AGN \citep{CMNS-1991}. \citet{JM97}
expli\-citly use the 99 per cent CL for magnitude
variations with \textit{amplitudes} exceeding
$2.576\,\sigma$,\footnote{Though we know that the value
  $2.576\,\sigma$ corresponds to 99.5 per cent (see below).} assuming
a normal distribution.  In \citet{1999A&AS..135..477R}, an explicit
definition for $C$ is given (equation~\ref{eq:C}), where the amplitude of
the target$-$comparison DLC has been changed by its dispersion, in an
attempt to compensate for the extreme sensibility of the \citet{JM97}
criterion to systematic (mostly type-I) errors (the practical reason
for this choice is illustrated in Section~\ref{sec:discu}).  Thus, the
parameter $C$ is the result of trying to improve the estimation of the
data errors, providing a variability criterion as strong as possible
against false positives arising from systematic errors.
 
However, we saw above that the $C$ criterion gives different results
than the $F$ test. Since the $F$ test is firmly rooted in a
statistical theoretical background, whereas the $C$ is a rather
loosely grounded criterion (that eventually got to be considered as an
actual test), we decided to carefully analyse the latter.

Putting aside for the moment the particular case of comparing light
curves, in a general setup the goal of both the $C$ and the $F$
statistics is to compare the dispersions ($C$ criterion) or variances
($F$ test) of two samples, taken from unknown populations. Both carry
out the comparison by rejecting (or not) the null hypothesis that both
dispersions and variances are statistically the same. Let
$C={\sigma_1}/{\sigma_2}$, and $F={\sigma_1^2}/{\sigma_2^2}$, where
$\sigma_1$ and $\sigma_2$ are the dispersions being compared, with
$\sigma_1 > \sigma_2$ in the case of the $F$ statistic. We discard
here any explicit scaling factor, because we are not computing results
of the tests but comparing them, so the numerical values of the
dispersions are irrelevant here.

\begin{figure}
\centering
\includegraphics[width=0.48\textwidth]{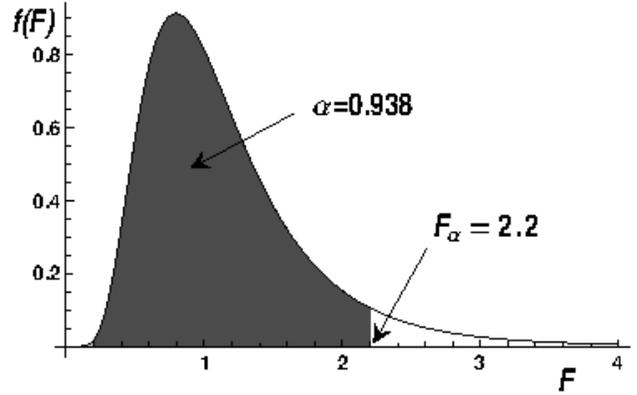}
\caption{Example of a Fisher $F$ density distribution, here with
  $\nu_1=20$, $\nu_2=15$, i.e. the sample with the larger dispersion
  has 21 measurements, and the other one 16. The CL is
  chosen here as $\alpha=0.938$, which gives a value of
  $F_\alpha=2.2$. If it turns out that $F_{\rm obs}>F_\alpha$, the
  null hypothesis is rejected; otherwise, the null hypothesis is not
  rejected.}
\label{fig:fisher}
\end{figure}

In order to make a theoretically based comparison between the methods,
we recall here the procedure for the $F$ test. First, we have to choose a
CL $\alpha$, that is, the complement of the probability
that two variances will give by chance an $F$ value so large that the
null hypothesis should be rejected. If, for example, one chooses 1
per cent as the above$-$mentioned probability, then
$\alpha=0.99$. Secondly, the \lq degrees of freedom\rq~$\nu_i=n_i-1,
i=1,2$ are computed, where $n_i, i=1,2$ are the number of measurements
of each sample. Thirdly, by using the probability density distribution
of the statistical variable $F$ with $\nu_1$ and $\nu_2$ degrees of
freedom, a value $F_\alpha$ is found, such that the area below the
distribution mentioned before to the left of $F_\alpha$ be $\alpha$
(Fig.~\ref{fig:fisher}). Fourthly, a value $F_{\rm
  obs}=\sigma_1^2/\sigma_2^2$ is computed from the measurements, by
using for each sample the usual formula

\begin{equation}
\sigma^2={1 \over n-1}\sum_{i=1}^n (x_i-\mu)^2,
\label{sigma}
\end{equation}

\noindent where $n$ is the size of the sample, $x_i$ are the measurements, and
$\mu$ is the mean of the sample, i.e., the sum of the measurements
divided by $n$.  Finally, $F_{\rm obs}$ is compared against
$F_\alpha$. If $F_{\rm obs}>F_\alpha$, then the null hypothesis is
rejected; otherwise, the null hypothesis is not rejected.

In turn, for the case of $C$ we have: first, the value $C_{\rm obs}$
is computed from the measurements, using the square root of
equation (\ref{sigma}) for each sample. Secondly, this value is (always)
compared with the number 2.576, irrespective of the number of
measurements. If $C>2.576$, the null hypothesis is rejected at a fixed
99.5 per cent CL.

So, the $C$ \lq test\rq~is not properly a statistical test. Tracing
back the origin of the fixed numbers 2.576 and 99.5 per cent, it seems
that they come from a standard {\it rejection of a bad measurement
  procedure}. According to this, given a set of measurements of a
given quantity, we can always compute the variance of the sample by
means of equation (\ref{sigma}).  Under the hypotheses that the
measurements came with a Gaussian distribution of errors, and that the
mean and the dispersion of the sample are good estimators of the true
mean and dispersion of the population of measurements, one might
discard those measurements that fall far enough from the mean of the
sample because those measurements can be regarded highly improbable
(some ins\-trumental or operational error rather than to an error by
chance). How far they should be from the mean in order to be discarded
depends on the experiment; usually, this distance is measured in units
of the dispersion of the sample. If this distance is taken as
$1\sigma$, for instance, it is said that the measurement is rejected
at a 68 per cent CL, because the area below a Gaussian
inside the abscissae $x=\pm\sigma$ is approximately 0.68.  But we may
invert the argument and put forward a CL, finding what
is the abscissa that gives that area.  If one chooses, for example,
0.995 as the level, then one obtains $x=\pm 2.576\,\sigma$ ($C$
critical value).

In this way, $C$ is not a strict, theoretically supported statistical
estimator
\footnote{Appendix~\ref{apendice} describes a possible implementation
  of a statistical test based on the ratio of dispersions of two
  distributions.}. As we have seen, the rejection of a bad measurement
works by comparing a given measurement with the \textit{mean of the
  distribution density of the measurements}, and measuring the
distance to that mean in terms of the \textit{dispersion of the
  distribution density of the measurements}. In the $C$ criterion,
however, a dispersion $\sigma_1$ is compared with a reference
dispersion $\sigma_2$, as if this last value were the mean of the
distribution density of dispersions, and the ratio $\sigma_1/\sigma_2$
becomes the distance, as if $\sigma_2$ were also the dispersion of the
distribution density of dispersions. That is, for the $C$ criterion to
work, $\sigma_2$ should be both the mean and the dispersion of the
(unknown) distribution of dispersions. And, it should be pointed out
that, whereas $C$ is strictly positive, and clearly the domain of a
distribution density of dispersions is the set of positive reals plus
zero, the $C$ criterion assumes a Gaussian distribution of
dispersions, i.e., a domain equal to the set of all real numbers.

\section{Results for field stars}
\label{sec:discu}

To better understand the results presented in Section~\ref{sec:stsv},
we analysed the stability of the statistics using the field stars. To
perform this, we considered all the selected stars in the frames,
excluding the AGN, and we calculated the $C$ and $F$ parameters for
all the DLCs using the same comparison and control stars as in the
case of the corres\-ponding AGN. By {\it selected} stars, we mean
those (between 6 and 44 per field) making the set of candidates from
which the compa\-rison and control stars were finally chosen. We
removed from this sample DLCs that were affected
by saturation, cosmic rays, stars that were too close to the edge of
the frames and any other evident defect. DLCs with $\Delta m \ge
0.4$\,mag were also discarded; this should remove any remaining very
ill-behaving DLC as well as known variables \citep[e.g., star S in the
  field of 3C\,279, a known variable with amplitude
  $>1$\,mag;][]{RVL98}. The original number of DLCs was $1039$, and
after the cleaning process, we had 981 DLCs left for their study.

The first thing to note is that 16.9 per cent of the DLCs are found to
be variable with the $F$ test, while this percentage drops to 9.5 per
cent using the $C$ criterion (in both cases, the $\Gamma$ correction
was applied). It is known \citep[e.g.][and references
  therein]{CvBB-2011} that the fraction of variable stars in a given
survey is a function of the survey parameters $-$time span and
sampling of the observational series, photometric precision$-$, as
well as the magnitudes, spectral types and luminosity classes of the
stars. As a general guide, from ground-based data, \citet{H-2008} says
that only 7 per cent of the stars are expected to vary at a 0.01 mag
precision level. \citet{CvBB-2011}, in turn, present a detailed
variability analysis based on {\it Kepler} data, with a time
resolution $\sim 30$\,min. From their results, it can be inferred that
the fraction of stars in our AGN fields (mostly located at relatively
high Galactic latitudes) that vary at a level $> 0.01$\,mag within a
few hours should be almost negligible $-$at most, well below 10 per
cent.

It is clear that both criteria classify as \lq variable\rq~a
larger$-$than$-$expected number of DLCs. However, this is particularly
evident for the $F$ test: 76 out of 981 DLCs (7.7 per cent) change
form NVar with the $C$ criterion to Var using the $F$ test (the
converse holds for a negligible 0.3 per cent, i.e., just three DLCs, so we
do not discuss this Var/NVar case).  In order to further inquire into
the reasons for this behaviour, we again analysed the distribution of
the different parameters characterizing the DLCs, as was done for the
AGN light curves.  The general results are qualitatively similar to
those presented in Sections~\ref{sec:distri} and
\ref{sec:histos}. However, it is worth mentioning that the most
significant diffe\-rences between distributions (supported by the KS
test) correspond to the ratio between the variability amplitude
($\Delta m$) and the scaled rms of the control light curve
($\Gamma\sigma_2$). While DLCs in the NVar/NVar case cluster at
$\Delta m/(\Gamma\sigma_2) \la 9$, those in the Var/Var case have a
broad distribution from $\Delta m / (\Gamma\sigma_2) \ga 9$ upwards;
the NVar/Var case, in turn, shows a narrow distribution centred at
$\Delta m / (\Gamma\sigma_2) \simeq 9$. For the observed DLCs of the
AGN sample, we obtained a similar result regarding the behaviour of
the ratio $\Delta m/(\Gamma \sigma_2)$ (also supported by the KS
test).

This means that both parameters agree in their classification for
almost all DLCs displaying variations with amplitudes above $\sim 9
\,\Gamma\sigma_2$ (Var/Var), and for most DLCs with $\Delta m \la 9
\,\Gamma\sigma_2$ (NVar/NVar), while a minor fraction of DLCs lying
within a narrow range around the limiting value ($\Delta m \simeq 9
\,\Gamma\sigma_2$) are classified as variable by the $F$ test and
non-variable by the $C$ criterion. Thus, both parameters behave as
sort of \lq$\sigma$-clipping\rq~criteria, but with different clipping
factors.  In this regard, it must be noted that if we apply the
original criterion proposed by \citet{JM97}, i.e. $\Delta m >
2.576\,\Gamma\sigma$, more than half the field stars DLCs (52.4 per
cent) are classified as variable.  On the other hand, if no weighting
($\Gamma$ factor) is applied, 20.7 per cent and 33.4 per cent of the
stars are classified as variable with the $C$ criterion and $F$ test,
respectively. Clearly, results from unweighted tests would be
catastrophic, and we will no longer discuss them.

\begin{figure}
\centering
\includegraphics[width=0.48\textwidth]{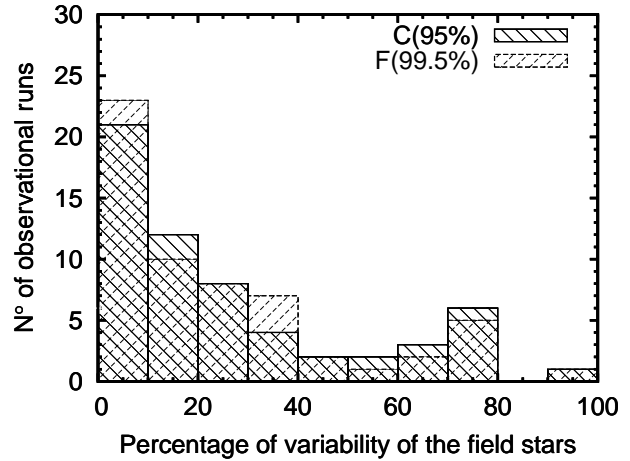}
\caption{Distribution of percentages of stars per field that resulted variable 
using $C$ at 95 per cent$-$CL and $F$ at 99.5 per cent$-$CL.}
\label{fig:starsCF}
\end{figure}

As a further comparison between different tests, we calculated the
percentage of DLCs in each star field that resulted to be varia\-ble
using the $C$ criterion and $F$ test, considering three different CLs:
95 per cent, 99 per cent, and 99.5 per cent. We found that the
distributions (for both statistics and the three CLs) have a clear
peak around 10 per cent, although, at the same CL, the histograms
corresponding to the $F$ test extend to larger variability
percentages.  It is interesting to note that the distributions of
$F_{99.5}$ and $C_{95}$, as shown in Fig.~\ref{fig:starsCF}, are
practically identical (a KS test gives a value of 1-prob$= 0.001$).
We interpret that, for our data, we have to relax the CL of the $C$
criterion to 95 per cent in order to obtain similar results as with
the $F$ test at the 99.5 per cent CL.

It is now clear that the $F$ test is not working as expected (and
neither does the $-$statistically ill founded$-$ $C$
criterion). However, this should not be surprising, since it is
well$-$known that the $F$ test is particularly sensible to non-Gaussian
errors \citep[e.g.][]{WJ-2012}, and photometric time series, unless
taken by an absolutely perfect space telescope equipped with an
absolutely perfect detector, will be affected by systematic error
sources, adding a \lq red-noise\rq~(i.e. time$-$correlated at low
frequencies) component. These sources of non-Gaussian distributed
errors include flat-field imperfections, airmass variations, imperfect
tracking, changing atmospheric conditions (seeing, transparency,
scintillation), chan\-ging moonlight and airglow illumination,
unnoticed cosmic rays, etc. Moreover, photometric errors usually
correlate with those systematic effects, as e.g. when the S/N ratio
drops due to changes in seeing or atmospheric transparency.

Any statistical test used to detect microvariability in AGN
DLCs obtained with ground-based telescopes should
thus be founded on solid theoretical bases and, at the same time, be
able to deal both with random (i.e., photometric) and
syste\-matic (non-Gaussian) errors. In a forthcoming paper, we will
further explore the performance of currently used tests by means of
simulated observations. This will allow us to test variability tests
under controlled situations, aiming at the selection of a test that is
appropriate to deal with real observational issues.

\section{Discussion}
\label{sec:summ}

There are several works that have been dedicated to the study of
statistical tools to detect microvariability in
AGN. \citet{2010AJ....139.1269D} studied the $\chi ^2$ test, the $F$
test for variances, the ANOVA test, and the $C$ criterion for a set of
simulated light curves, concluding that the most robust methodologies
are the ANOVA and $\chi ^2$ tests, while the $F$ statistic is less
powerful but still a reliable tool, and, finally, the $C$ criterion
should be avoided because it is not a proper statistical test. Further
analysis about these tests is presented in \citet{dD2014}, where a
study of the {\it Bartels and Runs} non-parametric test was added. In
that work, the author proposed that the best choices to detect
microvariability in AGN light curves are the use of an ANOVA or an
enhanced$-F$ test (in the latter, several comparison stars are used to
define a combined variance, instead of using a single star). A
continuation of this work was published by
\citet{2015AJ....150...44D}, where the enhanced-$F$ and the {\it
  nested} ANOVA tests were studied, concluding that these are the most
powerful tests to detect photometric variations in DLCs, due to the
increase in the power of the statistics, product of adding more
comparison stars to the statistical analysis (the {\it nested} ANOVA
test also requires some extra field stars, but fewer than in the
enhanced-$F$ test).

It should be noted that, in these papers, the authors explicitly
state that only photon shot$-$noise was considered for the light$-$curve
simulations, while any systematic effect was \lq entirely
disregarded\rq. So, despite their theoretical advantages, some of
these tests may be impractical for dealing with real observations;
moreover, if error distributions do not fulfil the assumptions on
which those tests are based, their use should be discouraged or, at
the very least, be taken with extreme care. In our case, we are
working with DLCs with a rather small number of
observations; this is a common situation, since AGN microvariability
light curves are mostly limited to under $\sim 30-40$ points
\citep[e.g.][]{KGC-2015}. The need of a large number of points in
light curves strongly limits the use of the $\chi^2$ test. The same
applies to the ANOVA test: despite its claimed power to detect
microvariabi\-lity \citep{2010AJ....139.1269D, dD2014}, this test is
seldom used, because it requires a large number of data points too
\citep{2011MNRAS.412.2717J}; moreover, data grouping might be
impractical for faint objects requiring relatively long integration
times, and could lead to false results if data within a time span
larger than the (unknown) variability time-scale are grouped.  In
fact, some doubtful results from the use of the ANOVA test in AGN
microvariability studies \citep{1998ApJ...501...69D} have already been
discussed in \citet{1999A&AS..135..477R}. Regarding the {\it nested}
ANOVA and the enhanced$-F$ tests, both tools require se\-veral
comparison stars to perform optimally \citep{2015AJ....150...44D},
while having appropriately populated star fields around AGNs is more
the exception than the rule. \citet*{2010ApJ...723..737V}, in turn,
discuss the application of different tests to AGN light curves from
space-based observations. They compare the $C$ criterion and the $\chi^2$ and
$F$ tests using a sample of randomly generated light curves, concluding
that the three tools show equal powers. However, when error
measurements are themselves erroneous, $\chi^2$ has the highest power
followed by $C$ and then $F$.

On the other hand, the use of tests specifically devised to deal with
Gaussian errors may not be optimal to work with ground-based light
curves, where atmospheric and instrumental effects produce correlated
errors, with non-Gaussian distributions. In fact, even under pure
random noise, errors in magnitude space will have asymmetric
non-Gaussian distributions \citep[e.g.][]{2010ApJ...723..737V}. This is
particularly relevant for the $\chi^2$ test, which requires that
individual data points have accurately determined errors, with
Gaussian distributions \citep[e.g.][]{2011MNRAS.412.2717J}; neither
of these is always fulfilled by optical ground-based photometry. The
$F$ test, in turn, does not behave as expected if error distributions
are non-Gaussian \citep[e.g.][]{WJ-2012}. It is thus important to
emphasize that $-$besides limitations typical of ground-based
observations$-$ variability stu\-dies of AGNs usually have particular
issues, like poorly sampled DLCs (due to low brightness of the
source), and the a\-vai\-la\-bi\-li\-ty of rather few field stars for
differential photometry; these facts must be taken into account for
the correct choice of the statistical analy\-sis of the DLCs.

\section{Summary and conclusions}
\label{sec:conc}

In order to test the most widely used tests for AGN variability, we
studied the $C$ and $F$ statistics with a large and homogeneous sample
of real observational data. We worked with a sample of 39 southern
AGNs observed with the 2.15m \lq Jorge Sahade\rq~telescope ({\sc
  CASLEO}), San Juan, Argentina, obtaining 78 nightly differential
photometry light curves, to which we applied the $C$ and $F$
statistics.

Besides which statistic is the better choice to analyse the behaviour
of the DLCs, we want to point out that it is very important to use the
weighted tests for the case of AGN differential photometry, because
of the particular issues mentioned in the previous paragraph
\citep[see also][for a full discussion on this
  issue]{2007MNRAS.374..357C}. We used the $\Gamma$ scale introduced
by \citet{1986PASP...98..802H}. There are cases in which the
variability results change just because of not using this
weight. Those cases are the ones in which $\Gamma$ is far from $1$
(i.e., the magnitudes of the comparison and/or control stars are not
similar to the target's magnitude).

From the results of applying the $C$ criterion and $F$ test to the
sample, we found that, with respect to the DLC amplitude ($\Delta m$),
$F$ results tend to classify as variable those DLCs with $\Delta m$
near the rms error, while for DLCs with high amplitude, both
statistics tend to detect variability. For the elapsed time ($\Delta
t$), DLCs with high values of $\Delta t$ are classified as variable,
in agreement to the fact that this high value usually appears in light
curves where systematic variability is observed. Both statistics seem
to be robust in the detection (or non-detection) of variability when
DLCs present low instrumental dispersion (0.012\,mag), but if the
dispersion of the \lq control$-$comparison\rq~light curve reaches
values larger than 0.02\,mag (some cases for the NVar/NVar histogram,
Fig.\ref{fig:hs2}c), low-amplitude AGN variability could be masked due
to the low S/N ratio in the DLC.

Taking a deeper look into the $C$ criterion, and comparing it with the
$F$ test, we arrived at the conclusion that, even though the $C$
criterion cannot be considered as an actual statistical test, it could
still be a useful parameter to detect variability, provided that the
correct significance factor is chosen. In this way, we found that
applying $C$ we may obtain rather more reliable variability results,
especially for small amplitude and/or noisy DLCs.

Finally, a study of the behaviour of the field stars was made in order
to analyse the stability of $C$ and $F$, excluding the AGN.  From
these new set of DLCs, we calculated the parameters involved in the
statistics and the percentage of field stars that result variable for
both $C$ and $F$. We found that, for the three CLs considered (95 per
cent, 99 per cent and 99.5 per cent), both statistics show a peak
around 10 per cent in their distributions, and comparing within the
same CL, the $F$ test presents an extended distribution to larger
va\-ria\-bi\-li\-ty percentages. We thus notice that the $F$ test tends to
classify as variable a larger number of DLCs than the $C$ parameter,
well above the expected number of variable stars in our fields. These
variability results are clearly false positive results, possibly due
to the inability of the $F$ test to deal with non-Gaussian distributed
errors.

There has to be always a balance between the power of a given test
(i.e. its ability to detect real variability) and its rate of false
po\-sitives. Ultimately, the outcome of this balance should be
dictated by astrophysical considerations, but this requires precise
know\-ledge of each test's behaviour under particular observational
conditions.

This study is being completed carrying out a series of si\-mu\-la\-ted
observations, which involve differential photometry for se\-veral AGNs
and comparison stars, immersed in a variety of distinct atmospheric
conditions and several different observational situations. Results
will be presented in a forthcoming paper.

\section*{Acknowledgements}

The present work was supported by the Argentine Agency ANPCyT (GRANT
PICT 2008/0627). LZ would like to thank the anonymous referee for the
useful comments. JAC, GER, IA, SAC and DC are CONICET researchers. DC
acknowledges financial support from PIP 0436 - CONICET, Argentina, and
from Proyecto G/127, UNLP, Argentina. GER has been supported by grant
AYA 2013-47447-C3-1-P (MINECO, Spain). JAC was supported on different
aspects of this work by Consejer\'{\i}a de Econom\'{\i}a,
Innovaci\'on, Ciencia y Empleo of Junta de Andaluc\'{\i}a under
excellence grant FQM-1343 and research group FQM-322, as well as FEDER
funds. This research has made use of the NASA/IPAC Extragalactic
Database (NED) which is operated by the Jet Propulsion Laboratory,
California Institute of Technology, under contract with the National
Aeronautics and Space Administration.

%\bibliographystyle{mn2e}
%\bibliography{biblio}

%%%%%%%%%%%%%%%%%%%%%%%%%%%%%%%%%%%%

\appendix

\section{The distribution density function of the $D$ statistic}
\label{apendice}

In order to determine whether two dispersions $\sigma_1$ and
$\sigma_2$ are not statistically equivalent, a statistical test should
be used. An equi\-valent test may be developed in which, instead of
the ratio of the variances as in the $F$ test, the ratio of the
dispersions is used, as in the $C$ parameter. In other words, we can
convert the $C$ statistic into a statistical test. This new test
should give no different results than the $F$ test. We will call it
the $D$ test. With this new statistic, one follows the same steps as
in the $F$ test: choosing a CL $\alpha$, computing the
value $D_\alpha$ that leaves an area $\alpha$ to its left below the
curve of the distribution density, finding the observed $D_{\rm
  obs}=\sigma_1/\sigma_2$ with $\sigma_1 > \sigma_2$, and rejecting
the null hypothesis if it happens that $D_{\rm obs} > D_\alpha$.

Suppose that, from a mother population with Gaussian pro\-bability
density and (unknown) dispersion $\sigma$, a series of samples of $n$
members each are taken. For each sample, its sample variance $s^2$ can
be computed as

\begin{equation}
s^2={1 \over n-1}\sum_{i=1}^n (x_i-\mu)^2,
\end{equation}

\noindent where $x_i$ is the $i-$th member of the sample, and

\begin{equation}
\mu={1\over n}\sum_{i=1}^n x_i
\end{equation}

\noindent is the mean of the sample. Hereafter, as a matter of convenience, we
will use the number of degrees of freedom $\nu=n-1$ instead of the
number of members $n$.  The sample variances $s^2$ of the diffe\-rent
samples have their own probability density distribution $f(s^2)$,
given by \citep{KS69}\footnote{ In \cite{KS69} the probability density
  distribution of $f(s^2)$ with its normalizing constant appears only
  in the specialized case $\sigma=1$ (their Eq.(11.41)).}

\begin{equation}
f(s^2\mid\nu,\sigma)=\left( \nu \over 2\right)^{\nu\over2} {(s^2)^{\nu/2-1}
\over \sigma^\nu \Gamma(\nu/2)} \exp\left(-{\nu \over 2}{s^2\over
\sigma^2}\right),
\end{equation}

\noindent which depends on the parameters $\nu$ and $\sigma$. Taking into
account that ${\rm d}(s^2)=2s\,{\rm d}s$, it is easy to find the
probability density distribution $g(s)$ of the sample dispersions $s$:

\begin{equation}
g(s\mid\nu,\sigma)={ \nu^{\nu\over2} \over 2^{\nu/2-1}} {s^{\nu-1} \over
\sigma^\nu \Gamma(\nu/2)} \exp\left(-{\nu \over 2}{s^2\over \sigma^2}\right),
\end{equation}

Now, given the distributions $g(s_1|\nu_1,\sigma_1)$ and
$g(s_2|\nu_2,\sigma_2)$ of the dispersions of two set of samples, each
with its own number of degrees of freedom $\nu_1$ and $\nu_2$, and
maybe taken from different mother populations with true dispersions
$\sigma_1$ and $\sigma_2$, one can find the distribution of their
quotient $D=s_1/s_2$ as \citep[][sect. 11.6]{KS69}

\setlength\arraycolsep{1.4pt}
\begin{eqnarray}
\lefteqn{h(D\mid\nu_1,\nu_2,\sigma_1,\sigma_2) =} \nonumber \\
&=&\int_0^\infty g(Dx\mid\nu_1+1,\sigma_1)
g(x\mid\nu_2+1,\sigma_2) x\, {\rm d}x \nonumber \\ 
&=&2{\Gamma({\nu_1+\nu_2 \over2})
\over \Gamma({\nu_1 \over 2})  \Gamma({\nu_2 \over 2})}
{\nu_1^{\nu_1/2} \nu_2^{\nu_2/2} \sigma_1^{\nu_2} \sigma_2^{\nu_1} 
D^{\nu_1-1} \over
(\nu_2 \sigma_1^2+ D^2 \nu_1 \sigma_2^2)^{(\nu_1+\nu_2)/2}}.
\end{eqnarray}

However, this result is completely useless because we do not know the
true dispersions $\sigma_1$ and $\sigma_2$. But, if the mother
population of both sets of samples is the same, or both sets come from
populations with the same dispersion (i.e. $\sigma_1=\sigma_2\equiv
\sigma$), then we have that the probability density distribution of
the ratio $D$ is

\begin{equation}
h(D\mid\nu_1,\nu_2) =
2{\Gamma({\nu_1+\nu_2 \over2}) \over \Gamma({\nu_1 \over 2}) 
\Gamma({\nu_2 \over 2})}
{\nu_1^{\nu_1/2} \nu_2^{\nu_2/2} D^{\nu_1-1} \over
(\nu_2 + D^2 \nu_1 )^{(\nu_1+\nu_2)/2}},
\label{Ddist}
\end{equation}

\noindent which is independent of the true dispersion. This turns out to be the
important point: this distribution is then ready to be used in a
statistical test. In particular, since it is the result of assuming
$\sigma_1=\sigma_2$, the $D$ test null hypothesis is that both $s_1$
and $s_2$ are statistically equivalent.

If the distribution equation (\ref{Ddist}) is compared with the well$-$known
distribution for the $F$ statistic, 

\begin{equation}
f(F\mid\nu_1,\nu_2) =
{\Gamma({\nu_1+\nu_2 \over2}) \over \Gamma({\nu_1 \over 2}) 
\Gamma({\nu_2 \over 2})}
{\nu_1^{\nu_1/2} \nu_2^{\nu_2/2} F^{\nu_1/2-1}\over
 (\nu_2 + F \nu_1 )^{(\nu_1+\nu_2)/2}},
\end{equation}

\noindent we see that they are the same distribution, only expressed with
different variables, i.e. $h(D){\rm d}D=f(F){\rm d}F$ with
$F=D^2$. Thus, using the $D$ test of dispersions gives exactly the
same results as using the $F$ test of variances.

% \bsp % ``This paper has been produced using the ...''

\label{lastpage}

\end{document}